\documentclass[journal=cgdefu,manuscript=article]{achemso}

\usepackage[version=3]{mhchem} % Formula subscripts using \ce{}
\usepackage{bm}% bold math
\usepackage{multicol}
\usepackage{siunitx}
\usepackage{graphicx}% Include figure files
\usepackage{dcolumn}% Align table columns on decimal point
\usepackage{soul}
\usepackage{lineno}
\mciteErrorOnUnknownfalse

\newcommand{\bsx}{\boldsymbol{x}}
\newcommand{\bsn}{\boldsymbol{n}}
\newcommand{\grad}{\nabla}
\def\div{\ensuremath{\nabla}\cdot}
\newcommand{\dd}{\mathrm{\,d}}

\def\rm#1{\mathrm{#1}}

\SectionNumbersOn

\author{Adrian Moure}
\email{amoure@caltech.edu}
\author{Xiaojing Fu}%
\email{rubyfu@caltech.edu}
\affiliation{%
 Department of Mechanical and Civil Engineering, California Institute of Technology, Pasadena, CA, United States \\
}%

\title{A phase-field model for wet snow metamorphism    }

\abbreviations{IR,NMR,UV}

\begin{document}

\begin{abstract}
The microstructure of snow determines its fundamental properties such as the mechanical strength, reflectivity, or the thermo-hydraulic properties.
Snow undergoes continuous microstructural changes due to local gradients in temperature, humidity or curvature, in a process known as snow metamorphism.
In this work, we focus on wet snow metamorphism, which occurs when temperature is close to the melting point and involves phase transitions amongst liquid water, water vapor, and solid ice. 
We propose a pore-scale phase-field model that simultaneously captures the three relevant phase-change phenomena: sublimation (deposition), evaporation (condensation), and melting (solidification).
The phase-field formulation allows one to track the temperature evolution amongst the three phases and the water vapor concentration in the air. 
Our three-phase model recovers the corresponding two-phase transition model when one phase is not present in the system. 2D simulations of the model unveils the impact of humidity and temperature on the dynamics of wet snow metamorphism at the pore scale. We also explore the role of liquid melt content in controlling the dynamics of snow metamorphism in contrast to the dry regime, before percolation onsets. 
The model can be readily extended to incorporate two-phase flow and may be the basis for investigating other problems involving water phase transitions in a vapor-solid-liquid system such as airplane icing or thermal spray coating.
\end{abstract}


\section{Introduction}
Snow and firn are heterogenous porous material composed of two components, water and air, distributed amongst three phases: solid ice, liquid water, and air with water vapor. The microstructure of snow and firn dictates their mechanical strength, reflectivity, and thermo-hydraulic properties, which govern important processes such as snow avalanches, snowpack and glacial hydrology, radar remote sensing, and the performance of snow vehicles \cite{colbeck1973theory}.
However, these properties constantly evolve because snow is a thermodynamically active material that undergoes continuous microstructural changes caused by phase transitions between ice, liquid water, and vapor, in a process known as snow metamorphism \cite{colbeck1996basic}. Ultimately, snow metamorphism leads to a denser snow composed of coarser and rounder ice grains. Based on whether liquid water is present, snow metamorphism is considered in two types ---dry and wet. Colbeck and other authors published a series of pioneer papers in the 70's and 80's \cite{colbeck1973theory,colbeck1980thermodynamics,colbeck1996basic,colbeck1982overview,colbeck1986statistics,colbeck1997review,raymond1979grain,wakahama1975role} that derived the theoretical foundations to understand the mechanisms driving each type of snow metamorphism.
Dry snow metamorphism \cite{colbeck1980thermodynamics,gundlach2018sintering} results from sublimation and deposition and is driven by the transport of water vapor between regions with different vapor pressure, i.e., regions with different temperature and/or grain curvature, according to the Gibbs-Thomson condition \cite{johnson1965generalization}. 
On the other side, wet snow metamorphism occurs when the temperature is close to the melting point. The presence of liquid water fundamentally alters the mechanisms of metamorphism by increasing the thermodynamic activity of snow and strengthening the mechanical and thermal connection between ice grains through capillary bridging \cite{colbeck1973theory,colbeck1980thermodynamics}.
Thus, compared to dry metamorphism, wet snow metamorphism displays accelerated coarsening and densification.
Colbeck proposed two different regimes for wet metamorphism depending on the liquid water content: the pendular and the funicular regimes \cite{colbeck1973theory}. 
In the funicular regime, the liquid water content is high so that the water completely surrounds the ice grains.
In the pendular regime, the liquid water content is low and most of the ice surface is in contact with air.
Thus, the pendular regime displays lower heat flow, grain growth, and snow densification, and larger capillary forces and grain-to-grain mechanical strength, compared to the funicular regime.
In both cases, vapor kinetics plays an important role \cite{colbeck1996basic}. While Colbeck's work provided mechanistic understanding of snow metamorphism,  it focused on theoretical snow configurations under ideal environmental conditions.
Subsequent experimental works used X-ray microtomography (microCT) to study the evolution of the dry snow microstructure in realistic scenarios under isothermal  conditions \cite{lowe2011interfacial,kaempfer2007observation} and temperature gradients \cite{pinzer2012vapor,wang2014evolution}. Many authors leveraged the microCT data to perform a comprehensive analysis of local crystal growth laws \cite{fife2014dynamics,krol2016analysis} and validate mean field models \cite{legagneux2005mean,krol2018upscaling}.
The first experiments of wet metamorphism focused on fully saturated snow \cite{colbeck1986statistics}.
Improved experimental techniques enable studies of wet metamorphism on samples with low liquid water content \cite{brun1989investigation,ebner2020liquid,stein1997monitoring,coleou1998irreducible}. These works analyzed snow wetness and grain size and derived empirical laws for the snow properties under wet conditions \cite{brun1992numerical}. However, more advanced techniques are still being developed to robustly visualize liquid water within snow samples, such as the use of hyperspectral imaging to map out liquid water content in snow \citep{donahue2022}.

%%%%%%%%%%%%%%%%%%%%%%%%%%%%%
Mathematical models may complement experimental development to provide insights into the wet snow metamorphism process. To this end, wet snow is modeled as a three-phase system composed of liquid water, vapor and solid ice, and thus can be grouped with other class problems such as snowflake growth and water drop icing in a humidity-controlled atmosphere. In this broader group of literature, some have proposed simplified 1D model to describe the role of humidity on the evolution dynamics \cite{sebilleau2021air}.
More elaborated 2D continuum models resorted to the phase-field method \cite{hagiwara2017ice,zhang2022phase} or the level-set method \cite{vahab2016adaptive,vahab2022fluid}, which have been successful in reproducing the phase transitions between two phases of water, i.e., solidification \cite{cristini2004three,wang1993thermodynamically,karma1996phase}, sublimation \cite{bouvet2022snow,kaempfer2009phase,reitzle2019direct}, and evaporation \cite{fei2022droplet,schweigler2017experimental,ronsin2020phase,wang2021phase}. However, most of these continuum models are limited to the two-phase regime, and do not simultaneously capture the transitions amongst the three water phases; thus, these models do not address the wet snow metamorphism problem.

%%%%%%%%%%%%%%%%%%%%%%%%%%%%%%%%
Here, we propose a phase-field model that describes the non-equilibrium evolution of the three-phase mixture composed of liquid water, water vapor and solid ice.
The modeling choice is inspired by recent applications of phase-field methods in porous media problems with phase transitions, including the modeling of microstructural evolution of porous materials and multicomponent fluid mixtures \cite{kaempfer2009phase,fu2016, shimizutanaka2017,mao2019,BHOPALAM2022115507,Yue_2020}, phase transitions during two-phase flow in porous media\cite{fu2017,cueto-felgueroso2018} and crystallization flow in porous media \cite{fu2018,fujimenez2020,zhang2022desal}. The three-phase model proposed in this work reduces to the classic two-phase model in the absence of a third phase, and the resulting two-phase models readily capture the Gibbs-Thomson conditions. 
The model allows us to analyze the pore-scale dynamics of that dictates wet snow metamorphism.  Currently, the model does not account for granular compaction, melt film imbibition or melt percolation. Rather, the primary focus of the model is to elucidate the role of thermodynamic-driven phase transitions on microstructure evolution of a quiescent wet snow mixture. 

\section{A phase-field model for wet snow mixture }

The phase-field method \cite{gomez2019review,boyer2006study} is a mathematical technique that is based on a free energy description of multi-phase mixtures, and is well-suited for problems with moving interfaces. Within the phase-field framework, one defines a dimensionless \textit{phase field} variable, denoted $\phi_l \in [0,1]$, which represents the volume fraction of the phase $l$ at any given point in the domain. The phase variable smoothly transitions across the interface between two distinct phases (Fig.~\ref{fig:1}). With the appropriate \textit{evolution equations}, the moving boundary problem is reformulated in a fixed domain, which avoids most of the numerical issues caused by the moving interfaces.

Here, we propose a \textit{non-variational} phase-field model that captures the phase transitions between ice, liquid water, and water vapor, i.e., solidification, sublimation, evaporation, and the opposite transitions. For simplicity, from here on we denote the liquid water as water and the water vapor as vapor. We assume that the air is at the atmospheric pressure. We assume that ice and water have equal density for the phase evolution equations, which introduces small mass conservation error. Although fluid flow is not considered in the current model, future work that incorporates fluid flow and unequal density may mitigate the mass conservation issue.

We note that in the spirit of numerical robustness and physical consistency, a  \textit{variational} phase-field formulation \cite{steinbach_2009} is often more desirable to ensure non-negative entropy production and the convergence towards equilibrium \cite{folch_plapp2005,toth2015}. However, due to the close proximity in the density of the ice and liquid water phase, we find that formulating a free energy for the wet snow system based on the density field introduces numerical singularities in the formulation. 

%The model leverages the phase-field formulation to track the temperature evolution amongst the three phases and the water vapor concentration in the air.This allows us to simultaneously capture the actual kinetics of the three relevant phase-change phenomena: sublimation (deposition), evaporation (condensation), and melting (solidification).
%
%The model also accounts for the nucleation of liquid water and ice, which permits to reproduce the transition between two- and three-phase systems.

\subsection{Model variables}
\begin{figure}[ht!]
\centering
\includegraphics[width=0.5\linewidth]{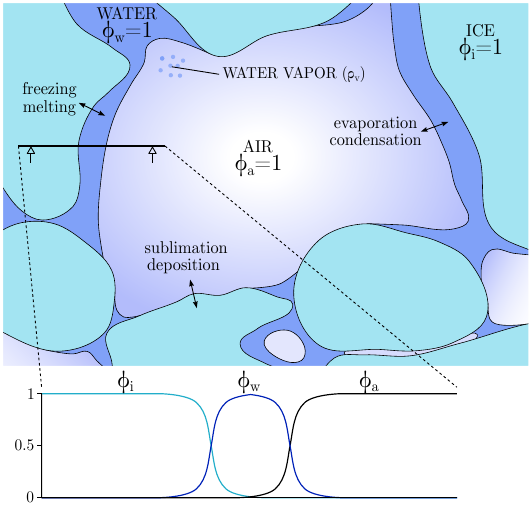}
\caption{Illustration of the key model variables $\phi_a, \phi_w, \phi_i,\text{ and } \rho_v$ as defined in the continuum domain composed of wet snow. The 1D transect profile illustrates the diffusive nature of the phase variables at the phase-phase boundaries. The model variable temperature $T$ (not shown here) is also defined point-wise in the entire domain.
} 
\label{fig:1}
\end{figure}

Our model unknowns are the phase field variables for ice $\phi_i(\bsx,t)$, water $\phi_w(\bsx,t)$, and air $\phi_a(\bsx,t)$ phases (see Fig.~\ref{fig:1}), as well as variables for the temperature $T(\bsx,t)$ (units $^\circ$C) and the vapor density $\rho_v(\bsx,t)$ (units kg/m$^3$). All variables are defined point-wise in the problem domain $\Omega$ (Fig.~\ref{fig:1}). 
Note that even though temperature and vapor density are not phase field variables, here we leverage the formulation to specify phase-dependent temperature evolution (see Section \ref{sec:energy_equation}), and to localize the vapor dynamics onto the air phase (see Section~\ref{sec:vapor_equation}). 

\subsection{Model equations}

In this section, we will introduce the five partial differential equations that describe the co-evolution of our model variables.

\subsubsection{The phase evolution equations}

The basic form of the phase evolution equations for the three phase field variables are expressed as:
\begin{align} 
     &\frac{\partial \phi_i}{\partial t} =  -\frac{M_0}{\Sigma_i} \frac{\delta \mathcal{F}^\rm{tri}(\overline{\phi})}{\delta \phi_i}  - \alpha_\rm{sol}\phi_i^2\phi_w^2\frac{T-T_\rm{melt}}{L_\rm{sol}/c_{p,w}} + \alpha_\rm{sub}\phi_i^2\phi_a^2\frac{\rho_v-\rho_{vs}^I(T)}{\rho_i} +\theta_\rm{nucl}(\overline{\phi},T), \label{eq:ice_simpl}\\ 
     &\frac{\partial \phi_w}{\partial t} = -\frac{M_0}{\Sigma_w} \frac{\delta \mathcal{F}^\rm{tri}(\overline{\phi})}{\delta \phi_w} + \alpha_\rm{sol}\phi_i^2\phi_w^2\frac{T-T_\rm{melt}}{L_\rm{sol}/c_{p,w}} +  \alpha_\rm{eva}\phi_w^2\phi_a^2\frac{\rho_v-\rho_{vs}^W(T)}{\rho_w} -\theta_\rm{nucl}(\overline{\phi},T), \label{eq:wat_simpl}\\  
     &\frac{\partial \phi_a}{\partial t} = -\frac{M_0}{\Sigma_a} \frac{\delta \mathcal{F}^\rm{tri}(\overline{\phi})}{\delta \phi_a} - \alpha_\rm{sub}\phi_i^2\phi_a^2\frac{\rho_v-\rho_{vs}^I(T)}{\rho_i} - \alpha_\rm{eva}\phi_w^2\phi_a^2\frac{\rho_v-\rho_{vs}^W(T)}{\rho_w}, \label{eq:air_simpl}
\end{align}
where $\displaystyle \delta\cdot/\delta \cdot$ denotes the variational derivative and we use the notation $\overline{\phi}=\{\phi_i,\phi_w,\phi_a\}$. Note that Eqs.\eqref{eq:ice_simpl}-\eqref{eq:air_simpl} account for total mass conservation by assuming equal density of ice and water ($\rho_i=\rho_w$).
%
%Eqs.~\eqref{eq:ice_simpl} to \eqref{eq:air_simpl} are the phase evolution equations. 
The first terms in the right-hand side of these three equations capture the Allen-Cahn kinetics of phase-field variables \cite{allen1979microscopic}, where $M_0$ is the phase-dependent mobility defined below in Sect.~\ref{sec:phase_dependent_funct} and  $\mathcal{F}^\rm{tri}$ is the classical energy functional of a ternary mixture \cite{boyer2006study}:
\begin{equation}\label{eq:energy_funct}
    \mathcal{F}^\rm{tri}(\overline{\phi})=\int\left( \frac{3}{\varepsilon} F^\rm{tri}(\overline{\phi}) + \frac{3}{2}\varepsilon\left[\Sigma_i(\grad\phi_i)^2 +\Sigma_w(\grad\phi_w)^2 +\Sigma_a(\grad\phi_a)^2 \right] +\beta(1-\phi_i-\phi_w-\phi_a) \right)\dd\Omega.
\end{equation}
The parameter $\varepsilon$ represents the width of the phase-field interface and $\beta$ is the Lagrange multiplier (see Section \ref{sec:final_equations}). The function $F^\rm{tri}$ is the \textit{triple-well potential}, defined as: 
\begin{equation}\label{eq:Ftri}
    F^\rm{tri}(\overline{\phi})=\frac{1}{2}\Sigma_i\phi_i^2(1-\phi_i)^2+\frac{1}{2}\Sigma_w\phi_w^2(1-\phi_w)^2+\frac{1}{2}\Sigma_a\phi_a^2(1-\phi_a)^2 +\Lambda\phi_i^2\phi_w^2\phi_a^2,
\end{equation}
where the $\Lambda$-term improves the dynamic consistency of the model\cite{boyer2006study} to mitigate spurious effect along phase-phase boundaries. 
The parameters $\Sigma_i$, $\Sigma_w$, and $\Sigma_a$ are related to the surface tensions ($\sigma_{lm}$) between phases $l$ and $m$ such that 
\begin{equation}\label{eq:surf_tens_eta}
    \Sigma_i = \sigma_{iw} +\sigma_{ia} -\sigma_{wa},\quad 
    \Sigma_w = \sigma_{iw} +\sigma_{wa} -\sigma_{ia}, \quad
    \Sigma_a = \sigma_{wa} +\sigma_{ia} -\sigma_{iw}.
\end{equation}
%
%\begin{align}\label{eq:surf_tens_eta}
%    \Sigma_i = \sigma_{iw} +\sigma_{ia} -\sigma_{wa},\\
%    \Sigma_w = \sigma_{iw} +\sigma_{wa} -\sigma_{ia},\\
%    \Sigma_a = \sigma_{wa} +\sigma_{ia} -\sigma_{iw}.
%\end{align}
%
We extend the Allen-Cahn kinetics with the $\alpha$-terms in Eqs.~\eqref{eq:ice_simpl} to \eqref{eq:air_simpl}, which account for the different phase transitions. In particular, the $\alpha_\text{sol}$ term accounts for solidification or melting, the $\alpha_\text{sub}$ term accounts for sublimation or deposition, and the $\alpha_\text{eva}$ term accounts for evaporation or condensation. 
The estimation of $\alpha$'s parameter values is described in the Supporting Information (Section~\ref{sec:2PMequivalence} and Table~\ref{tab:kinetic_constr}).
The parameters $L_\rm{sol}$, $c_{p,w}$, $\rho_w$ and $\rho_i$ are the solidification latent heat, water specific heat capacity, water density, and ice density, respectively.
We follow \citet{kaempfer2009phase} and express the saturated vapor density of ice $(\rho_{vs}^I)$ and water $(\rho_{vs}^W)$ as
\begin{equation}\label{eq:rhovsJ}
    \rho_{vs}^J(T)=\rho_a 0.62\frac{P_{vs}^J(T)}{P_a-P_{vs}^J(T)},
\end{equation}
where the index $J$ stands for $I$ and $W$, $\rho_a$ is the dry air density, $P_a$ is the atmospheric pressure (units Pa), and $P_{vs}^J$ is the saturated vapor pressure of phase $J$ (units Pa). The saturated vapor pressure of ice and water are defined as \cite{flatau1992polynomial,wexler1977vapor}
\begin{align} \label{eq:PJvsI}
    &P_{vs}^I(T)=\rm{exp}\left( \sum_{j=0}^4 k_j(T+T_0)^{j-1} + k_5\,\rm{ln}(T+T_0)\right),\\ \label{eq:PJvsW}
    &P_{vs}^W(T)=\rm{exp}\left( \sum_{j=0}^6 g_j(T+T_0)^{j-2} + g_7\,\rm{ln}(T+T_0)\right),
\end{align}
respectively, where $T_0=273.15\SI{}{\celsius}$, and the values of the parameters $k_j$ and $g_j$ are listed in Table~\ref{tab:sat_vapor} in the Supporting Information.
% Note that we take $T_0=273.15$ and the units of $T$ are $\SI{}{\celsius}$.

%%%%%%%%%%%%%%%%%%%%%%%%%%%%%
The $\theta_\rm{nucl}$ terms in Eqs.~\eqref{eq:ice_simpl}--\eqref{eq:wat_simpl} account for ice and water nucleation, which are necessary in order to trigger the nucleation of a third phase when only two phases exist initially. In particular, here we account for the nucleation of water when $T$ increases above $T_\rm{melt}$ in an ice-air system and the nucleation of ice when $T$ decreases below the ice nucleation temperature $T_\rm{nucl}$ in a water-air system.
We propose a phenomenological nucleation term
\begin{equation}\label{eq:nucleation}
    \theta_\rm{nucl}(\overline{\phi},T) = \alpha_\rm{nucl}M_0\phi_a^2 [\phi_w^2 N_i(T)-\phi_i^2 N_w(T)],
\end{equation}
where $\alpha_\rm{nucl}$ is a parameter controlling the strength of the nucleation and the ice and water nucleation functions $N_i(T)$ and $N_w(T)$, respectively, are expressed as
\begin{equation}\label{eq:nucl_terms}
    N_i(T) = 0.5-0.5\,\rm{tanh}[20(T-T_\rm{nucl}+0.1)], \quad  N_w(T) = 0.5+0.5\,\rm{tanh}[20(T-T_\rm{melt}-0.1)].
\end{equation}
Note that $N_i$ becomes active when $T<T_\rm{nucl}$, which allows the presence of undercooled water in the range $T_\rm{nucl}<T<T_\rm{melt}$.
%
%such that $N_i>0$ if $T<T_\rm{nucl}$ and $N_i\approx 0$ if $T>=T_\rm{nucl}$, which allows the presence of undercooled water in the temperature range $T_\rm{nucl}<T<T_\rm{melt}$. 
%Analogously, $N_w>0$ if $T>T_\rm{melt}$ and $N_i\approx 0$ if $T<=T_\rm{melt}$.
%
According to Eq.~\eqref{eq:nucleation}, ice nucleation occurs on the water-air  interface when $T<T_\rm{nucl}$ (note the term $\phi_a^2\phi_w^2$ multiplying $N_i$).
Once ice nucleation advances, the water-air interface splits into the water-ice and ice-air interfaces, resulting in $\phi_a^2\phi_w^2=0$ and thus $\theta_\rm{nucl}$ stays zero even if $T<T_\rm{nucl}$.
Water nucleation exhibits an analogous behavior.
Note that $\theta_\rm{nucl}$ only needs to be active until the system enters into the spinodal decomposition region of the ternary mixture.

%%%%%%%%%%%%%%%%%%%%%%%%%%%%%

\subsubsection{Energy conservation equation} \label{sec:energy_equation}

Here, we formulate conservation of energy in terms of temperature, expressed as
\begin{equation}
   \rho(\overline{\phi})c_p(\overline{\phi})\frac{\partial T}{\partial t} = \div\left[ K(\overline{\phi})\grad T \right] + \rho(\overline{\phi})L_\rm{sol}\left(\frac{\partial \phi_a}{\partial t} + \frac{\partial \phi_i}{\partial t}\right) -\rho(\overline{\phi})L_\rm{sub}\frac{\partial \phi_a}{\partial t}. \label{eq:tem_simpl}
\end{equation}
Eq.~\eqref{eq:tem_simpl} accounts for thermal diffusion and the latent heat released and/or absorbed during the different phase transitions. $L_\rm{sub}$ is the sublimation latent heat. The latent heat of evaporation is then captured through $L_\rm{eva}=L_\rm{sub}-L_\rm{sol}$. The phase-dependent density ($\rho$), specific heat capacity ($c_p$), and thermal conductivity ($K$) are defined as:
\begin{equation}
    \rho(\overline{\phi})=\sum_{l=i,w,a}\phi_l\rho_l;\;\;\; 
    c_p(\overline{\phi})=\sum_{l=i,w,a}\phi_l c_{p,l};\;\;\;
    K(\overline{\phi})=\sum_{l=i,w,a}\phi_l K_l, \label{eq:property_interp}
\end{equation}
where $\rho_l$, $c_{p,l}$, and $K_l$ are the density, specific heat capacity, and thermal conductivity, respectively, of phase $l$. We assume that $\rho_l$, $c_{p,l}$, and $K_l$ are constant in time. Note that here we violate our primary assumption that $\rho_i=\rho_w$ used in Eqs.~\eqref{eq:ice_simpl}-\eqref{eq:air_simpl}, which incurs mass conservation error; instead, we consider different density for the ice and water phase for the energy equation in order to adopt the correct heat capacity for each phase. Although doing so does not eliminate errors in energy conservation due to errors in mass conservation, it improves the thermal consistency of the model.

%\begin{align}
%    \rho(\overline{\phi})&=\phi_i\rho_i+\phi_w \rho_w+\phi_a \rho_a,\label{eq:rho_interp}\\ 
%    c_p(\overline{\phi})&=\phi_i c_{p,i}+\phi_w c_{p,w}+\phi_a c_{p,a},\label{eq:cp_interp}\\ 
%    K(\overline{\phi})&=\phi_i K_i+\phi_w K_w+\phi_a K_a, \label{eq:K_interp}
%\end{align}
%

%%%%%%%%%%%%%%%%%%%%%%%%%%%%%%

\subsubsection{Mass conservation of vapor} \label{sec:vapor_equation}
The vapor mass conservation equation is expressed as 
\begin{equation}
    \frac{\partial(\phi_a\rho_v)}{\partial t}=\div\left[ \phi_a D_v(T)\grad \rho_v \right] + \rho_\rm{SE} \frac{\partial \phi_a}{\partial t},\label{eq:vap_simpl}
\end{equation}
which accounts for vapor diffusion in the air and mass transfer of water during sublimation and evaporation (and the opposite transitions). 
Here, we formulate a phase-dependent density $\rho_\rm{SE}(\overline{\phi})$ such that $\rho_\rm{SE}=\rho_i$ in case of sublimation and $\rho_\rm{SE}=\rho_w$ in case of evaporation. The specific form for $\rho_\rm{SE}(\overline{\phi})$ is discussed and provided in Eq.~\eqref{eq:rho_SE} of Section~\ref{sec:phase_dependent_funct}. 
$D_v$ is the vapor diffusion coefficient, whose dependence on $T$ (in $\SI{}{\celsius}$) follows the expression: \cite{massman1998review}

\begin{equation}\label{eq:vapor_diff}
    D_v(T)=D_{v0}\left(\frac{T+T_0}{T_0}\right)^{1.81},
\end{equation}
where $D_{v0}$ is the vapor diffusion coefficient at the freezing point and $T_0=273.15$.

%%%%%%%%%%%%%%%%%%%%%%%%%%%%%%

\subsection{Phase-dependent functions} \label{sec:phase_dependent_funct}
The model equations introduced above rely on a few parameters that are phase-dependent. For material property parameters such as $\rho, c_p, K$, we simply use phase-weighted averages as shown in Eq.~\eqref{eq:property_interp}. However, parameters such as the mobility function $M_0$ in Eqs.~\eqref{eq:ice_simpl}--\eqref{eq:air_simpl} and $\rho_\rm{SE}$ in Eq.~\eqref{eq:vap_simpl} are related to phase change dynamics at phase-phase boundaries or triple junction regions, and thus require more careful mathematical formulation. Here we resort to the ternary diagram to illustrate these modeling choices.

A ternary diagram is the common tool to represent three-phase systems. 
In a ternary diagram, the vertex $l$ represents the point $\phi_l=1$, while the opposite side of that vertex represents the points $\phi_l=0$ (for $l=\{i,w,a\}$). The interior of the diagram represents points where the three phases coexist. Thus, any valid $\{\phi_i,\phi_w,\phi_a\}$ configuration is represented by a point in the ternary diagram. In particular, evaporation takes place on the side $\phi_i=0$, sublimation on the side $\phi_w=0$, and solidification on the side $\phi_a=0$.

At the boundaries of the ternary diagram (i.e. $\phi_a=0$, $\phi_w=0$, and $\phi_i=0$), the mobility function $M_0$ must take the values $M_\rm{sol}$, $M_\rm{sub}$, and $M_\rm{eva}$ that correspond to the kinetics of solidification, sublimation and evaporation, respectively. The exact expressions for these parameters are provided in Eqs.~\eqref{eq:GT_sol1}, \eqref{eq:GT_sub1}, and \eqref{eq:GT_evap1} of the Supporting Information. 
Within the interior region of the ternary diagram, however, $M_0$ needs to be interpolated based on $M_\rm{sol}$, $M_\rm{sub}$, and $M_\rm{eva}$. A common choice is the cubic interpolation, which presents issues for two reasons\cite{miyoshi2016validation,moelans2022new}: (1) the triple junction is a region, not a point, where the three phases coexist; (2) the values of $M_\rm{sol}$, $M_\rm{sub}$, and $M_\rm{eva}$ range several orders of magnitude in this region (see Table~\ref{tab:kinetic_constr} in the Supporting Information).
Thus, a cubic interpolation of $M_0$ would result in large gradients in mobility within the interior of the ternary diagram, leading to severe numerical challenges. 
In addition, a non-unique value of $M_0$ within the triple junction is not physically consistent and result in unrealistic behavior of the triple junction.

To avoid these two issues, we propose a piecewise constant interpolation of $M_0$, similar to that proposed in \citet{miyoshi2016validation}, where $M_0$ takes a constant value near each of the three sides and in the interior region of the ternary diagram:
\begin{equation}\label{eq:mobility}
    M_0(\overline{\phi})=\left\{\begin{array}{ll}
    M_\rm{sol} &\rm{if}\; \phi_a<=\phi_i,\;\; \phi_a<=\phi_w, \;\;\rm{and}\;\; \phi_a < 0.01, \\
    M_\rm{eva} &\rm{if}\; \phi_i<\phi_a,\;\; \phi_i<=\phi_w, \;\;\rm{and}\;\; \phi_i < 0.01, \\
    M_\rm{sub} &\rm{if}\; \phi_w<\phi_a,\;\; \phi_w<\phi_i, \;\;\rm{and}\;\; \phi_w < 0.01, \\
    M_\rm{av} &\rm{otherwise},
    \end{array} \right.
\end{equation}
where we take $M_\rm{av}$ as the geometric mean (i.e., $M_\rm{av}=\sqrt[3]{M_\rm{sub}M_\rm{sol}M_\rm{eva}}$). We find that Eq.~\eqref{eq:mobility}, compared to a cubic interpolation, improves the numerical efficiency and allows us to simulate triple junctions while respecting the actual kinetics of the different phase transitions. 

The density $\rho_\rm{SE}$ must take the value $\rho_i$ in case of sublimation and $\rho_w$ in case of evaporation (i.e., $\phi_w=0$ and  $\phi_i=0$, respectively, in the ternary diagram). The value of $\rho_\rm{SE}$ along the side $\phi_a=0$ is irrelevant because $\rho_\rm{SE}\frac{\partial\phi_a}{\partial t}$ (see Eq.~\eqref{eq:vap_simpl}) is zero on that side. We propose a cubic interpolation between the sides $\phi_w=0$ and $\phi_i=0$ in the ternary diagram, such that the derivatives of $\rho_\rm{SE}(\overline{\phi})$ are zero on the sides $\phi_w=0$ and $\phi_i=0$. We define the function $\rho_\rm{SE}$ as
\begin{equation} \label{eq:rho_SE}
    \rho_\rm{SE}(\overline{\phi})=\left\{\begin{array}{ll}
    \frac{\rho_i+\rho_w}{2}+\frac{\rho_w-\rho_i}{4}x_\rm{SE}(3-x_\rm{SE}^2)  &\rm{if}\; \phi_a\neq 1, \\
    \frac{\rho_i+\rho_w}{2}  &\rm{if}\; \phi_a = 1, 
    \end{array} \right. 
    \quad \rm{with} \quad x_\rm{SE}=\frac{1-\phi_a-2\phi_i}{1-\phi_a}.
\end{equation}
Equation~\eqref{eq:rho_SE} displays a discontinuity at $\phi_a=1$. From a physical and numerical point of view, this discontinuity does not represent an issue because the term $\rho_\rm{SE}\frac{\partial\phi_a}{\partial t}$ is  zero if $\phi_a=1$.
Nevertheless, we implement a regularization of Eq.~\eqref{eq:rho_SE} (see Section~\ref{sec:discretization} in the Supporting Information).

\subsection{The Gibbs-Thomson condition}
The Gibbs-Thomson effect describes the deviations in the equilibrium chemical potential at a two-phase boundary due to the curvature of the interface. Such effect readily applies to the three types of interface considered in wet snow and influences both the kinetics and equilibrium conditions experienced by curved interfaces. 
In Section \ref{sec:2PMequivalence} of the Supporting Information, we show that our model is equivalent to the two-phase models for solidification, sublimation, and evaporation when only two phases are present, and the resulting reduced model equations recover the Gibbs-Thomson for two-phase boundaries under certain conditions. In particular, our model reproduces, under certain parameter regimes, the kinetics defined by the Gibbs-Thomson condition for ice-liquid interfaces:
\begin{equation}\label{eq:GTSOL}
    \frac{T-T_\rm{melt}}{L_\rm{sol}/c_{p,w}}=-d_\rm{sol}\chi-\beta_\rm{sol}v_n,
\end{equation}
where $T$ is the interface temperature, $\beta_\rm{sol}$ is the kinetic attachment coefficient, $d_\rm{sol}$ is the capillary length, $\chi$ is the curvature of the interface (positive for spherical ice grains), and $v_n$ is the normal velocity of the interface (positive for ice growth).
Note that, at thermodynamic equilibrium (i.e., $v_n=0$), Eq.~\eqref{eq:GTSOL} simplifies to the classic Gibbs-Thomson equation:
\begin{equation}\label{eq:Tequil}
T^\rm{equil}=T_\rm{melt}-\frac{L_\rm{sol}d_\rm{sol}}{c_{p,w}}\chi,
\end{equation}
which states that the actual equilibrium temperature for solidification is lower than $T_\rm{melt}=0^\circ$C for spherical ice grains. Similarly, our model captures, under certain parameter regimes, the kinetics defined by the Gibbs-Thomson condition for air-ice interfaces:
\begin{equation}\label{eq:GTSUB}
    \frac{\rho_v-\rho_{vs}^I(T)}{\rho_{vs}^I(T)}=d_\rm{sub}\chi+\beta_\rm{sub}v_n,
\end{equation}
and for air-liquid interfaces:
\begin{equation}\label{eq:GTEVP}
    \frac{\rho_v-\rho_{vs}^W(T)}{\rho_{vs}^W(T)}=d_\rm{eva}\chi+\beta_\rm{eva}v_n,
\end{equation}
where $\beta_\rm{sub}$ and $d_\rm{sub}$ are the Gibbs-Thomson coefficients for sublimation and $\beta_\rm{eva}$ and $d_\rm{eva}$ are the Gibbs-Thomson coefficients for evaporation. The model's versatility in capturing the Gibbs-Thomson conditions for all three types of two-phase boundaries rely on the phase-dependent formulation of $M_0$ as discussed in Section \ref{sec:phase_dependent_funct}.

Note that because we do not assume equal $\Sigma_\alpha's$, spurious phase (nonphysical third phase) may appear along phase-phase boundaries, leading to inaccuracies in the equilibrium interface solution. We mitigate this issue using the $\Lambda$-term in the free energy formulation (Eq.~\eqref{eq:Ftri}). In addition, we note that the Gibbs-Thomson condition is only recovered in phase-field models under certain parameter conditions \cite{kaempfer2009phase} or mathematical constructions through asymptotic analysis \cite{karma1998quantitative, karma2001, folch_plapp2005}. Our current formulation does not entail the mathematical additions needed to recover the Gibbs-Thomson condition for all interfaces for all parameter regimes; however, under certain parameter regimes, the model recovers these conditions. More details are provided in Section \ref{sec:kinet_param} and Supplementary Information.

\section{Numerical implementation} \label{sec:final_equations}
We first apply the phase constraint $\phi_w=1-\phi_i-\phi_a$ to Eqs.~\eqref{eq:ice_simpl}--\eqref{eq:air_simpl}, which allows us to compute the Lagrange multiplier $\beta$ in Eq.~\eqref{eq:energy_funct} and eliminate one phase evolution equation:
\begin{equation}
    \beta = \frac{1}{\frac{1}{\Sigma_a}+\frac{1}{\Sigma_w}+\frac{1}{\Sigma_i}}\frac{3}{\varepsilon}\left( \frac{1}{\Sigma_i}\frac{\partial F^\rm{tri}}{\partial\phi_i} + \frac{1}{\Sigma_w}\frac{\partial F^\rm{tri}}{\partial\phi_w} + \frac{1}{\Sigma_a}\frac{\partial F^\rm{tri}}{\partial\phi_a} \right)
\end{equation}
The resulting system of four coupled partial differential equations are:
\begin{align} 
     \frac{\partial \phi_i}{\partial t}&\quad  =&&   - \frac{3 M_0(\overline{\phi})}{\varepsilon\Sigma_T} \left[ (\Sigma_w+\Sigma_a)\frac{\partial F^\rm{tri}}{\partial \phi_i} -\Sigma_a\frac{\partial F^\rm{tri}}{\partial \phi_w} - \Sigma_w\frac{\partial F^\rm{tri}}{\partial \phi_a} \right] + 3M_0(\overline{\phi})\varepsilon \nabla^2\phi_i \nonumber   \\ 
     & \quad && - \alpha_\rm{sol}\phi_i^2\phi_w^2\frac{T-T_\rm{melt}}{L_\rm{sol}/c_{p,w}} + \alpha_\rm{sub}\phi_i^2\phi_a^2\frac{\rho_v-\rho_{vs}^I(T)}{\rho_i} \nonumber \\
     & \quad  &&+\alpha_\rm{nucl}M_0(\overline{\phi})\phi_a^2 [\phi_w^2 N_i(T)-\phi_i^2 N_w(T)],\label{eq:ice_final}\\
     \frac{\partial \phi_a}{\partial t}&\quad =&&  - \frac{3 M_0(\overline{\phi})}{\varepsilon\Sigma_T} \left[ -\Sigma_w\frac{\partial F^\rm{tri}}{\partial \phi_i} -\Sigma_i\frac{\partial F^\rm{tri}}{\partial \phi_w} + (\Sigma_i+\Sigma_w)\frac{\partial F^\rm{tri}}{\partial \phi_a} \right] + 3M_0(\overline{\phi})\varepsilon \nabla^2\phi_a \nonumber\\ 
     & \quad    && - \alpha_\rm{sub}\phi_i^2\phi_a^2\frac{\rho_v-\rho_{vs}^I(T)}{\rho_i} - \alpha_\rm{eva}\phi_w^2\phi_a^2\frac{\rho_v-\rho_{vs}^W(T)}{\rho_w}, \label{eq:air_final} \\  
   \frac{1}{\xi_T}\rho(\overline{\phi})c_p(\overline{\phi})\frac{\partial T}{\partial t}&\quad =&& \div\left[ K(\overline{\phi})\grad T \right] + \rho(\overline{\phi})L_\rm{sol}\left(\frac{\partial \phi_a}{\partial t} + \frac{\partial \phi_i}{\partial t}\right) -\rho(\overline{\phi})L_\rm{sub}\frac{\partial \phi_a}{\partial t},\label{eq:tem_final} \\ 
    \frac{1}{\xi_v}\frac{\partial(\phi_a\rho_v)}{\partial t}&\quad =&&\div\left[ \phi_a D_v(T)\grad \rho_v \right] + \rho_\rm{SE}(\overline{\phi}) \frac{\partial \phi_a}{\partial t}, \label{eq:vap_final}
\end{align}
where $F^\rm{tri}$, $\rho_\rm{SE}$ and $M_0$ are defined in Eqs.~\eqref{eq:Ftri}, \eqref{eq:rho_SE}, and \eqref{eq:mobility}, respectively. 
The saturated vapor densities $\rho^I_{vs}$ and $\rho^W_{vs}$ are defined in Eqs.~\eqref{eq:rhovsJ}--\eqref{eq:PJvsW}.
The parameter $\Sigma_T=\Sigma_i\Sigma_w+\Sigma_i\Sigma_a+\Sigma_w\Sigma_a$, where $\Sigma_l$'s are defined in Eq.~\eqref{eq:surf_tens_eta}.
The nucleation functions $N_i$ and $N_w$ are defined in Eq.~\eqref{eq:nucl_terms}.
The phase-dependent functions $\rho$, $c_p$, and $K$ are defined through Eq.~\eqref{eq:property_interp} and $D_v$ follows Eq.~\eqref{eq:vapor_diff}. The time scaling parameters $\xi_T$ and $\xi_v$ are introduced in the next section (Section \ref{sec:time_scaling}).
The rest of the model parameters are listed in Tables~\ref{tab:kinetic_constr} and \ref{tab:parameter_val} in the Supporting Information.

\subsection{Temporal scaling}\label{sec:time_scaling}
Wet snow metamorphism involves multiple coupled processes at different time scales. Assuming a characteristic length scale of $L=10^{-4}\,$m, the characteristic times for different processes can range from $\sim 10^{-4}\,\rm{s}$ for vapor diffusion to $\sim 10^{4}\,\rm{s}$ for sublimation (see Section~\ref{sec:SI_timescale} in the Supporting Information for detailed calculations). 
A monolithic numerical solver requires the use of time steps that resolve the fastest process (i.e., vapor diffusion at $\sim 10^{-4}\,\rm{s}$ ), which would significantly increase the computational cost. Here, we leverage the procedure explained in \citet{kaempfer2009phase} to speed up the simulations. This approach leverages the fact that $T$ and $\rho_v$ are quasi-steady compared to the phase transition kinetics. 
The procedure consists of multiplying the right-hand side of the $T$ and $\rho_v$ equations (Eqs.~\eqref{eq:tem_simpl} and \eqref{eq:vap_simpl}, respectively) by a time-scale factor ($\xi_T$ and $\xi_v$ respectively), so that the process of vapor and thermal diffusion is numerically slowed down by orders of magnitude while still maintaining a quasi-steady state in $T$ and $\rho_v$. 
This approach enables the use of larger time steps without introducing noticeable errors in the simulations. In this work, we take $\xi_v=10^{-3}$ and $\xi_T=1$ or $\xi_T=10^{-2}$ depending on whether or not solidification occurs in any parts of the domain.
More details on this procedure are described in Section \ref{sec:SI_timescale} of the Supporting Information.

\subsection{Kinetic parameters constraints}\label{sec:kinet_param}
In order to accurately capture the interface kinetics set by the Gibbs-Thomson conditions, certain model parameters need to be constrained by the interface physics. To this end, the relations between the Gibbs-Thomson parameters $(\beta_j,d_j)$ and our model parameters $(M_j,\alpha_j)$ (for $j=\{\rm{sol},\rm{sub},\rm{eva}\}$; see Section~\ref{sec:2PMequivalence}) can be formulated from asymptotic analysis \cite{karma1996phase,karma1998quantitative} but are only valid under certain conditions. In particular, we note that these asymptotic analysis do not readily apply to the wet snow problem as they are originally developed for compositional problems with symmetric \cite{karma1996phase, karma1998quantitative} and asymmetric diffusivities \cite{karma2001,ohno2016}. Solutions using anti-trapping current have been devised for the thermal problems, but only with symmetric diffusivities \cite{folch_plapp2005}. Thus, in order to robustly capture the Gibbs-Thomson condition under all parameter regimes for the wet snow problem, further mathematical additions remain to be developed.

We note, however, in the earlier and similar work on dry snow metamorphism by \citet{kaempfer2009phase}, it is shown that if we impose $\varepsilon<D\beta'$, where $D$ are the different thermal or vapor diffusivities and $\beta'$ is the scaled kinetic coefficient, then no anti-trapping current is needed to correctly recover the sharp-interface limit. In particular, for the wet snow problem, this corresponds to the following sets of constraints: 
\begin{align}
&\varepsilon < \beta_\text{sol}\frac{K_w}{\rho_w c_{p,w}}\approx 10^{-5} \text{m}; \;(
\text{for water-ice interface})\\ 
&\varepsilon < \beta_\text{sub}\frac{\rho_{vs}}{\rho_i}\frac{K_a}{\rho_a c_{p,a}}\approx 10^{-7} \text{m}; \;(
\text{for air-ice interface})\\
&\varepsilon < \beta_\text{eva}\frac{\rho_{vs}}{\rho_i}\frac{K_w}{\rho_w c_{p,w}}\approx 10^{-8} \text{m}. \;(
\text{for air-water interface})
\end{align}
These conditions impose a restriction on $\varepsilon$, which represents the diffusive interface width at phase-phase boundaries. The value of $\varepsilon$ dictates the spatial discretization of our model, which must be fine enough to resolve the interface width with sufficient numerical elements. Given that the most restrictive conditions of the above requires $\varepsilon<10^{-8}\,\rm{m}$, this severely restricts the domain size we can simulate. In order to speed up the simulations, we consider a value $\varepsilon<10^{-6}\,\rm{m}$, which allows us to use a coarser spatial discretization. As a result, our simulation are less accurate in capturing interface kinetics along the air-ice and air-water interfaces. However, considering the phase change kinetics involving the air phase is slow compared to the solidification/melting processes along the water-ice interface, we suspect the overall error on \textit{wet} snow processes is small.

\subsection{Numerical methods}

We perform both 1D and 2D simulations of our model equations using the finite element method. In particular, we use a uniform mesh composed of bilinear (linear in 1D) basis functions. For the time integration, we use a semi-implicit algorithm based on the generalized-$\alpha$ method \cite{chung1993time,jansen2000generalized}, where we treat implicitly all the terms except the function $M_0$ and the parameter $\xi_T$, which are constant during each time step. We use the Newton-Raphson method with adaptive time stepping scheme to solve the resulting nonlinear system. To avoid singularities and numerical issues we regularize the functions $\rho$, $c_p$, $K$, and $\rho_\rm{SE}$ (see Section \ref{sec:discretization} in the Supporting Information).

We impose no-flux boundary conditions for the phase variables and Dirichlet boundary condition for the temperature.  The vapor boundary condition is either Dirichlet or no-flux depending on specific case studies. In most simulations, we initialize the vapor density with the saturated vapor density for ice at the prescribed temperature.

%For the regularization of $\rho_\rm{SE}$ we
%(1) implement an additional cubic interpolation locally around $\phi_a=0$ which smooths out the discontinuity at that point, and
%(2) extend $\rho_\rm{SE}$ off the ternary diagram constant in the normal direction to the diagram sides.

\section{Results}
\subsection{Directional solidification and nucleation of a third phase}
\begin{figure}[ht!]
\centering
\includegraphics[width=\linewidth]{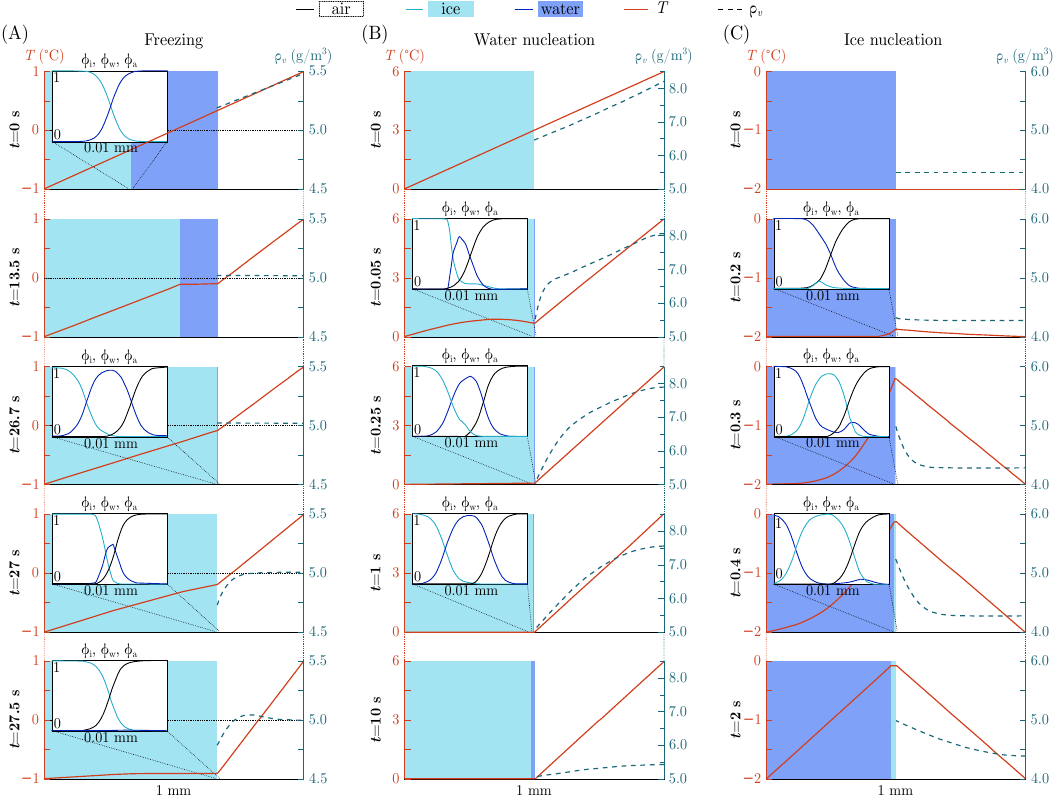}
\caption{Time evolution (from top to bottom) of (A) directional water freezing towards the vapor phase, (B) water nucleation during melting of an initially dry system, and (C) ice nucleation during an initially ice-free system. The shaded regions in the main panels represent the different phases (light blue, dark blue, and unshaded for ice, water, and air, respectively). The red solid line indicates the temperature (left axis) and the dashed blue line indicates the vapor density in the air (right axis). 
The insets show the diffusive profiles of $\phi_i$ (light blue), $\phi_w$ (dark blue), and $\phi_a$ (black) around the interface where phase change is occuring.
} 
\label{fig:2}
\end{figure}

In this section, we demonstrate the ability of our model to capture directional solidification as imposed by a temperature gradient, and the nucleation of a third phase in a system that is initially composed of two phases only. In particular, we present three 1D simulations in Fig.~\ref{fig:2} that show (a) the complete freezing of water in an initial ice-water-air system (Fig.~\ref{fig:2}A), (b) water nucleation in an initial ice-air system (Fig.~\ref{fig:2}B), and (c) ice nucleation in an initial water-air system (Fig.~\ref{fig:2}C).  To perform these 1D simulations, we impose fixed temperatures and no-flux boundary condition for the vapor phase on the two ends of the domain, which measures $\SI{1}{\milli\meter}$ in length and is discretized with 2000 elements.

In the freezing simulation (Fig.~\ref{fig:2}A), the initially imposed temperature gradient results in the directional solidification of water towards the air phase. At each time step, the temperature distribution appears piecewise linear, which confirms the quasi-steady behavior of $T$ as discussed in Section~\ref{sec:time_scaling}.
Although the $\rho_v$ dynamics are also quasi-steady, its distribution is not always linear (see $t=27$ and $\SI{27.5}{\second}$ in Fig.~\ref{fig:2}A). This nonlinear response is primarily triggered when a new ice-air interface is established upon complete freezing ($\sim \SI{27.5}{\second}$), and a new equilibrium is being established to satisfy the saturated vapor density in the presence of ice, prescribed by Eq.~\eqref{eq:PJvsI}.

%\red{[Right: the change is caused by (i) saturated vapor of ice and water are different and (ii) saturated vapor depends on temperature, which experiences a quick change when water phase vanishes (T=-0.1C to T=-1C from t=26.7 to t=27.5).]}
%This is because $\rho_v$ tends to the  saturated vapor concentration on the water-air and ice-air interfaces and $\grad\rho_v\cdot\bsn=0$ on the right edge of $\Omega$.
%

The nucleation simulations demonstrate significantly more nonlinear dynamics. In the water nucleation simulation (Fig.~\ref{fig:2}B), we impose $\SI{0}{\celsius}$ and $\SI{6}{\celsius}$ on the left and right boundaries of the domain respectively to induce melting at the initial ice-air interface. In the ice nucleation simulation (Fig.~\ref{fig:2}C), the water-air system is initially at $\SI{-2}{\celsius}$ uniformly and we consider the ice nucleation temperature $T_\rm{nucl}=\SI{-1}{\celsius}$. 
Ice nucleation onsets, preferentially towards the water phase, well before the temperature on the water-air interface increases to $T_\rm{nucl}$  (Fig.~\ref{fig:2}C, $t=\SI{0.2}{\second}$). At $t=\SI{0.3}{\second}$, when $T>T_\rm{nucl}$ on the interface, the ice phase has fully developed. 
Both types of nucleation described here are highly energetic events, leading to rapid changes in temperature and saturated vapor density at the nucleation region. 
As a result, the $T$ and $\rho_v$ profiles are no longer linear, in contrast to the directional freezing problem (Fig.~\ref{fig:2}A). 
We note that here, nucleation only occurs at the phase-phase interface due to its mathematical formulation in Eq.~\eqref{eq:nucleation}.

%\red{Response: It is normal that the $T$ and $\rho_v$ profiles are not linear during nucleation events! They are $\sim$highly energetic events (temperature quickly change at the nucleation region, idem for saturated vapor.
%The reason melting only occurs at the interface is the nucleation term $\theta_\rm{nucl}$, it only allows nucleation in the air-ice interface for melting, or the air-water interface for freezing.}

%%%%%%%%%%%%%%
We note that the phase field variables maintain the tanh-profile (quasi-equilibrium profile) during the freezing simulation (insets of Fig.~\ref{fig:2}A), but such profiles are temporarily lost during the nucleation simulations (insets Figs.~\ref{fig:2}B, C). 
In the latter cases, the loss of the tanh-profile is due to changes in mobility $M_0$ (Eq.~\eqref{eq:mobility}) during the dynamic nucleation events, and could result in small errors in curvature calculations and thus influence the kinetics of phase change through the Gibbs-Thomson condition. Nevertheless, the tanh-profile is recovered shortly after nucleation (e.g., at $t=\SI{1}{\second}$ in Figs.~\ref{fig:2}B, C); thus the undesired loss of the tanh-profile is temporary (occurs only during nucleation) and for this reason, we assume its impact on the overall kinetics of the problem is negligible.

\subsection{Simulation  of dry snow metamorphism in 2D}
\begin{figure}[t]
\centering
\includegraphics[width=\linewidth]{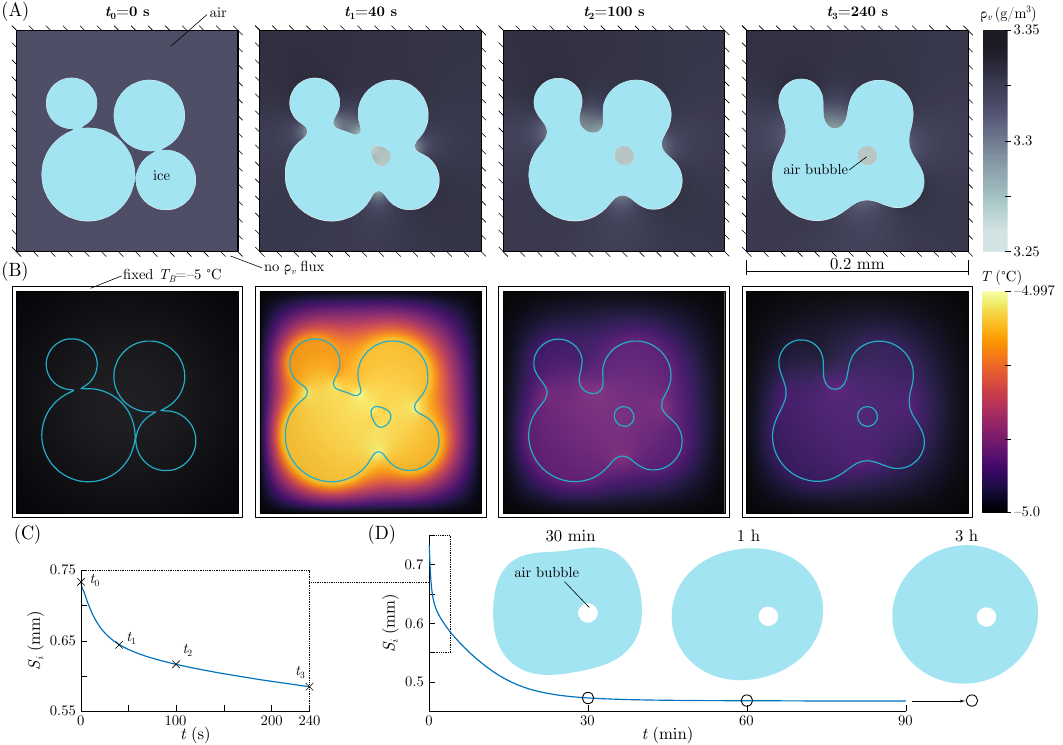}
\caption{Dry snow metamorphism.
Time evolution of (A) the ice geometry and vapor concentration and (B) the temperature distribution at $t_0=\SI{0}{\second}$, $t_1=\SI{40}{\second}$, $t_2=\SI{100}{\second}$, and $t_3=\SI{240}{\second}$.
The blue solid line in (B) is the isoline $\phi_i=0.5$.
Time evolution of the total ice perimeter $S_i$ for short (C) and longer (D) times. The ice geometry at $t=\SI{30}{\minute}$, $\SI{1}{\hour}$, and $\SI{3}{\hour}$ is included in (D).
} 
\label{fig:3}
\end{figure}

Under thermal conditions when liquid water phase is absent ($\phi_w$=0), our model reduces to the two-phase model for dry snow metamorphism  first proposed in \citet{kaempfer2009phase}:
\begin{align}\label{eq:cond2_sublice}
    \frac{\partial\phi_i}{\partial t} &= -3 M_0\left( \frac{1}{\varepsilon} \phi_i(1-\phi_i)(1-2\phi_i)   - \varepsilon \nabla^2\phi_i \right) + \alpha_\rm{sub}\phi_i^2\phi_a^2\frac{\rho_v-\rho_{vs}^I(T)}{\rho_i}\\
 \frac{1}{\xi_T}\rho(\overline{\phi})c_p(\overline{\phi})\frac{\partial T}{\partial t} &= \div\left[ K(\overline{\phi})\grad T \right]  -\rho(\overline{\phi})L_\rm{sub}\frac{\partial \phi_a}{\partial t},\\
    \frac{1}{\xi_v}\frac{\partial(\phi_a\rho_v)}{\partial t}&=\div\left[ \phi_a D_v(T)\grad \rho_v \right] + \rho_\rm{SE} \frac{\partial \phi_a}{\partial t},
\end{align}
with $\partial\phi_a/\partial t=-\partial\phi_i/\partial t$. 
Here, we show a 2D example of the dry metamorphism model using a simple snow geometry composed of four circular ice grains in contact (see $t_0$ in Fig.~\ref{fig:3}A).
The domain measures $0.2\times\SI{0.2}{\milli\meter^2}$ and is meshed with $400\times 400$ elements.
We impose a fixed temperature $T_B=\SI{-5}{\celsius}$ and no-flux of $\rho_v$ on all boundaries of the domain. Over the 90 minute time span of the simulation, we plot the time evolution of the ice phase $\phi_i$ and vapor concentration $\rho_v$ (Fig.~\ref{fig:3}A), along with the $T$ (Fig.~\ref{fig:3}B) and the total ice interface length $S_i$ (Figs.~\ref{fig:3}C and D).
Also known as coarsening, dry snow metamorphism under isothermal condition is a curvature-driven process, which minimizes the total interfacial length and the curvature of the system. The process penalizes regions of high curvature and is driven by vapor mass transfer between regions of different curvatures \cite{colbeck1980thermodynamics}.
Our simulation readily illustrates the above mechanism. We observe that the ice grains quickly sinter (before $t_1=\SI{40}{\second}$) and the coarsening process gradually slows down, as measured by the decay in slope in Figs.~\ref{fig:3}C and D. Accompanying these geometric changes are the dynamics in $T$ distribution due to latent heat generation in regions of higher curvature (see Fig.~\ref{fig:3}B). 
The simulation also captures the entrapment of an air bubble during the coarsening process.
At steady state (e.g., $t=\SI{3}{\hour}$ in Fig.~\ref{fig:3}D), the vapor concentration inside the bubble is set by its curvature according to the Gibbs-Thomson condition (Eq.~\eqref{eq:GTSUB}).

\subsection{Simulations of wet snow dynamics in 2D}

In this section, we perform a suite of numerical simulations of our full model to investigate wet snow dynamics as it undergoes phase transition. Due to high computational cost of the model, here we only focus on small sized domains ($\sim$mm) in 2D.

\subsubsection{Influence of vapor concentration}

\begin{figure}[ht!]
\centering
\includegraphics[width=\linewidth]{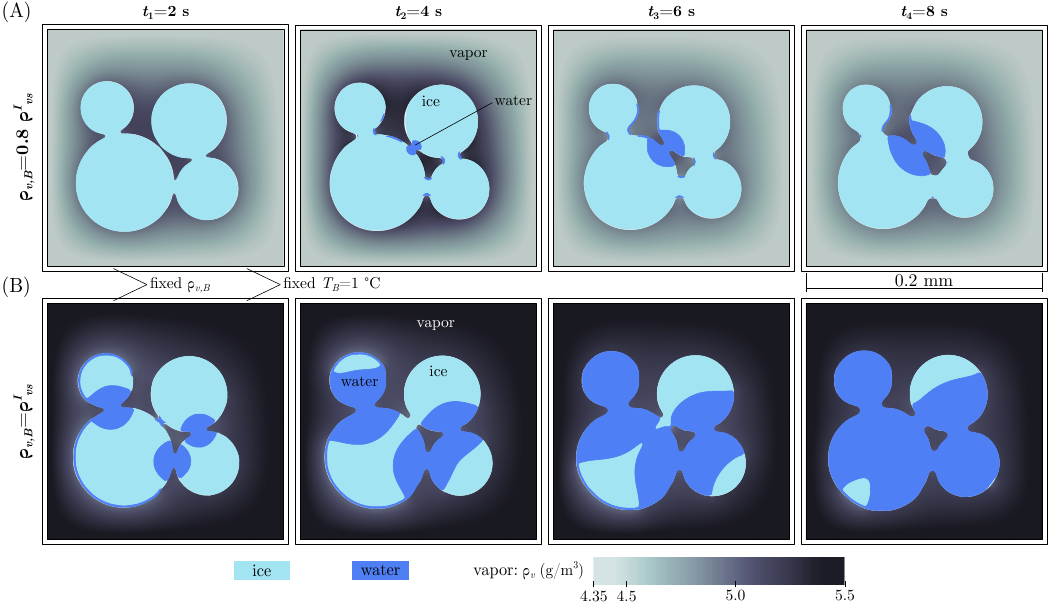}
\caption{Impact of vapor concentration on melting of snow.
Time evolution of the ice and water phases and vapor concentration for boundary vapor concentration of (A) $\rho_{v,B}=0.8\rho_{vs}^I$ and (B) $\rho_{v,B}=\rho_{vs}^I$ when snow melts with $T_B=\SI{1}{\celsius}$. The initial snow geometry is plotted in Fig.~\ref{fig:3}.
} 
\label{fig:4}
\end{figure}

In this part, we study the impact of vapor concentration on the kinetics of melt generation when the initially dry material is subjected to melting. To achieve this, we consider the same four-grain geometry as the initial dry snow simulation in Fig.~\ref{fig:3}, we impose a boundary temperature of $T_B=\SI{1}{\celsius}$ and run two simulations with different vapor concentrations on the boundary, denoted $\rho_{v,B}$. In particular, we impose an undersaturated vapor density, $\rho_{v,B}=0.8\rho^I_{vs}(\SI{1}{\celsius})$, in the first simulation (Fig.~\ref{fig:4}A) and a saturated vapor density, $\rho_{v,B}=\rho^I_{vs}(\SI{1}{\celsius})$, in the second one (Fig.~\ref{fig:4}B).

\begin{figure}[ht!]
\centering
\includegraphics[width=\linewidth]{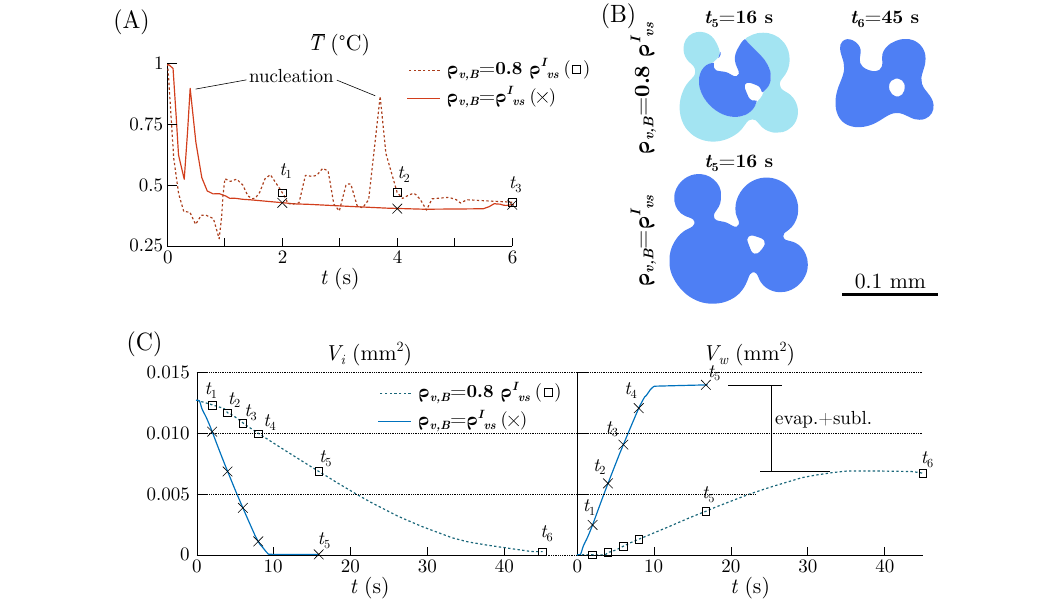}
\caption{More detailed analysis of the simulations shown in Figure \ref{fig:4}.
(A) Time evolution of the spatial-average temperature $(\overline{T})$ when $\rho_{v,B}=0.8\rho_{vs}^I$ (dashed line) and $\rho_{v,B}=\rho_{vs}^I$ (solid line).
(B) Ice and water phases at times $t_5=\SI{16}{\second}$ and $t_6=\SI{45}{\second}$ when $\rho_{v,B}=0.8\rho_{vs}^I$ (top) and at time $t_5=\SI{16}{\second}$ when $\rho_{v,B}=\rho_{vs}^I$ (bottom).
(C) Time evolution of the ice ($V_i$, left) and water ($V_w$, right) volume when $\rho_{v,B}=0.8\rho_{vs}^I$ (dashed lines) and $\rho_{v,B}=\rho_{vs}^I$ (solid lines).
} 
\label{fig:5}
\end{figure}

We quantify these two simulations by plotting the time evolution of the spatially averaged temperature $\overline{T}$, volume of ice phase $V_i$ and water phase $V_w$ in Figs.~\ref{fig:5}A and C. The results illustrate that, when the initial vapor phase is undersaturated, melt nucleation is delayed and the kinetics of melt generation is slower (Fig.~\ref{fig:5}C). We explain this from the perspectives of thermal balance. Because both sublimation and melt generation requires energy input, an unsaturated vapor phase will reduce the amount of energy available for melting. This is evident when we observe that the average temperature decrease is stronger in the unsaturated vapor case (Fig.~\ref{fig:5}A, dashed line). 
This lower temperature also causes a delay in water nucleation (Figs.~\ref{fig:4}A, Figs.~\ref{fig:5}C). We also compare the final snapshots of the two simulations, taken at different times ($t=\SI{45}{\second}$ for the first and $t=\SI{16}{\second}$ for the second simulation) but both correspond to a state when ice no longer exists (Fig.~\ref{fig:5}B). It is evident that there is significantly less melt water in the first simulation, which is caused by the significant sublimation that claims part of the ice in order to saturate the vapor phase.

Finally, we want to remark that while water nucleation occurs in many parts of the ice interface, melting into the ice phase only advances in some regions of the interface because the process is limited by the thermal energy influx (Figs.~\ref{fig:4} and \ref{fig:5}). In particular, we observe that melting usually starts at the grain contact regions because larger curvatures induce higher temperatures, as already shown in the case with dry snow (Fig.~\ref{fig:3}B).

%%%%%%%%%%%%%%%%%%%%%%%%%%%%%%%%
%%%%%%%%%%%%%%%%%%%%%%%%%%%%%%%%

\subsubsection{Influence of temperature}
\begin{figure}[t]
\centering
\includegraphics[width=\linewidth]{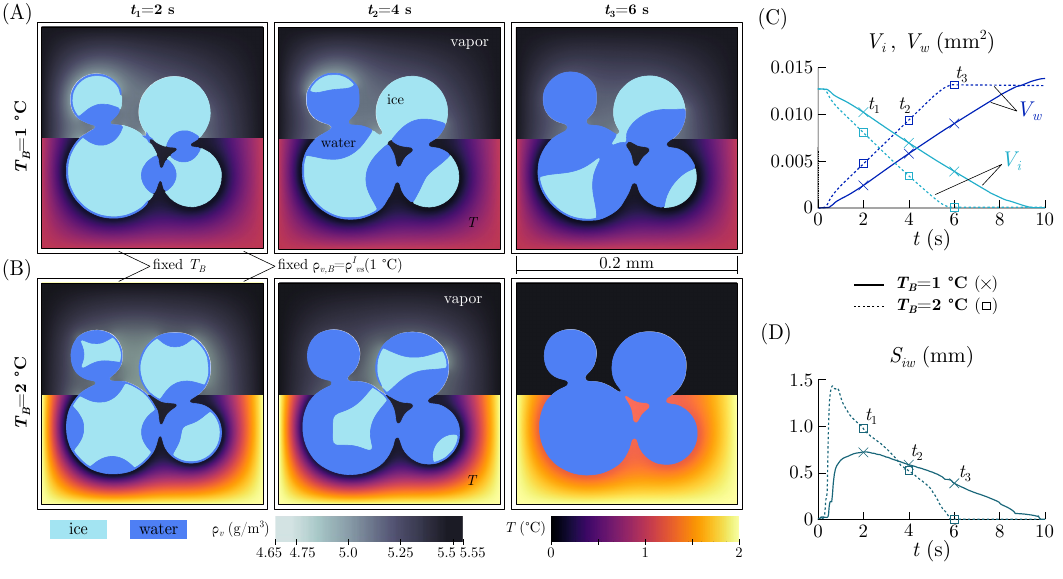}
\caption{Impact of temperature on melting of snow.
(A,B) Time evolution of the ice and water phases and vapor concentration (upper half of each panel) and temperature distribution (bottom half of each panel) for boundary temperature (A) $T_b=\SI{1}{\celsius}$ and (B) $T_b=\SI{2}{\celsius}$. The initial snow geometry is plotted in Fig.~\ref{fig:3}.
(C) Time evolution of the ice ($V_i$, light blue) and water ($V_w$, dark blue) volume for $T_b=\SI{1}{\celsius}$ (solid lines) and $T_b=\SI{2}{\celsius}$  (dashed lines).
(D) Time evolution of the ice-water interface length $S_{iw}$ for $T_b=\SI{1}{\celsius}$ (solid line) and $T_b=\SI{2}{\celsius}$  (dashed line).
} 
\label{fig:6}
\end{figure}

In this part, we study the impact of temperature on wet snow evolution. Using the same initial dry snow geometry (Fig.~\ref{fig:3}), here we fix the vapor concentration on the boundary such that $\rho_{v,B}=\rho^I_{vs}(\SI{1}{\celsius})$ and consider two different boundary temperature $T_B$'s, namely, $T_B=\SI{1}{\celsius}$ (Fig.~\ref{fig:6}A) and $T_B=\SI{2}{\celsius}$ (Fig.~\ref{fig:6}B).
We further quantify these two simulations with the time evolution of $V_i$ and $V_w$ (Fig.~\ref{fig:6}C) and the ice-water interface length $S_{iw}$ (Fig.~\ref{fig:6}D). As expected, we observe that melting is faster at a higher boundary temperature (Fig.~\ref{fig:6}C). 
In particular, our results illustrate that a higher temperature accelerates melting by promoting melt nucleation and thus increases the amount of ice-water interface available for melt advancing (Fig.~\ref{fig:6}D). In particular, we observe that a film of nucleated water completely surrounds the ice phase at the higher temperature (Fig.~\ref{fig:6}B), while water film only appears at some regions of the ice interface at the lower temperature(Fig.~\ref{fig:6}A). 
After nucleation, we observe that melting only advances at the three grain contact locations for low $T_B$, while melting progresses in a few more locations, some away from grain contacts, at high $T_B$ (Fig.~\ref{fig:6}B).

Finally, upon complete melting, the two simulations display a small difference in $V_w$ (Fig.~\ref{fig:6}C, at $t=\SI{10}{\second}$). This difference is caused by the imposed $T$ and $\rho_v$ boundary conditions. Since we impose $\rho_{v,B}=\rho^I_{vs}(\SI{1}{\celsius})$ for both simulations, vapor concentration on the boundary is out of equilibrium for the case of $T_B=\SI{2}{\celsius}$. This induces a vapor outflux through the boundary which is reflected in final value of $V_w$ via sublimation/evaporation. 

\subsubsection{Influence of liquid water content}

\begin{figure}[ht!]
\centering
\includegraphics[width=\linewidth]{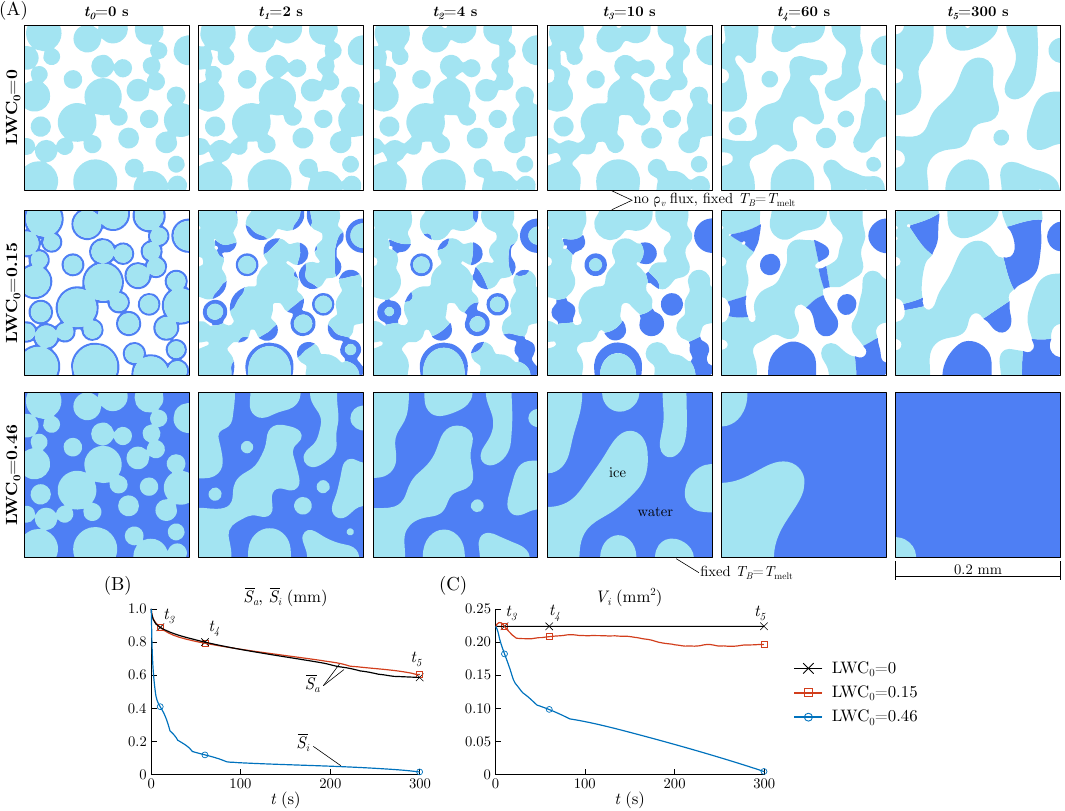}
\caption{Influence of LWC on quasi-isothermal snow metamorphism, where the boundary temperature $T_B$ is fixed to $T_\rm{melt}$.
(A) Time evolution of the ice and water phases for initial LWC (LWC$_0$) of 0 (top row), 0.15 (center row), and 0.46 (bottom row; fully wet).
(B) Time evolution of the normalized interface length of the air phase $(\overline{S}_a)$ for LWC$_0=0$ (black) and LWC$_0=0.15$ (red) and the ice phase $(\overline{S}_i)$ for LWC$_0=0.46$ (blue).
The interface length is normalized by its initial value.
(C) Time evolution of the ice volume $(V_i)$ for LWC$_0=0$ (black), 0.15 (red), and 0.46 (blue). 
} 
\label{fig:7}
\end{figure}

In this section, we study the influence of the liquid water content (LWC) on wet snow metamorphism. We consider a simplified snow geometry composed of circular ice grains in a domain of $0.2\times\SI{0.2}{\milli\meter^2}$ meshed with $800\times 800$ elements (Fig.~\ref{fig:7}A, $t=0$s).  We run three simulations with different initial LWC (LWC$_0$) that correspond to dry (LWC$_0=0$), partially wet (LWC$_0=0.15$), and fully wet (LWC$_0=0.46$) snow. The initial ice phase distribution is the same for all the three cases. For the partially wet snow, we impose a water film of $\SI{2.6}{\micro\meter}$ thickness uniformly surrounding the ice grains. For the fully wet snow, we fill the pore space entirely with the liquid phase. We adopt the parameter value $\varepsilon=2\times10^{-7}\,$m. To keep the system in quasi-isothermal conditions, we impose a fixed temperature $T_B=T_\rm{melt}$ and no-flux in terms of vapor density $\rho_v$ on all boundaries.

\begin{figure}[ht!]
\centering
\includegraphics[width=\linewidth]{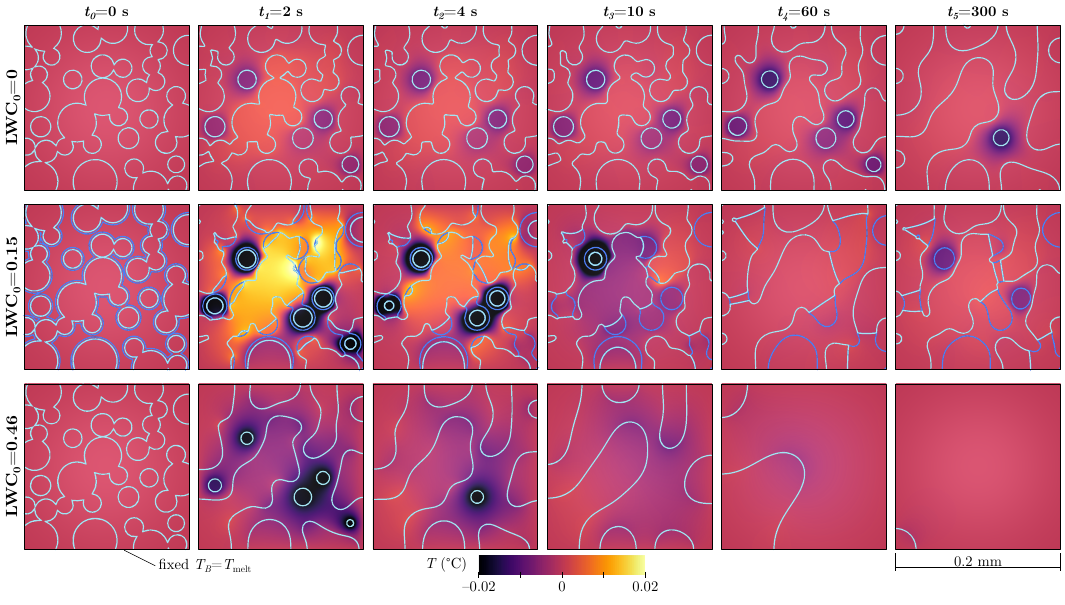}
\caption{Influence of LWC on quasi-isothermal snow metamorphism, where the boundary temperature $T_B$ is fixed to $T_\rm{melt}$. 
Time evolution of the temperature distribution for initial LWC (LWC$_0$) of 0 (top row), 0.15 (center row), and 0.46 (bottom row; fully wet).
Light and dark blue lines represent the ice and water interfaces, respectively.
} 
\label{fig:8}
\end{figure}

The simulations results (Figs.~\ref{fig:7} and \ref{fig:8}) illustrate the influence of LWC$_0$ on the detailed dynamics of metamorphism. 
In particular, when comparing against dry snow, we find that partially wet snow evolves towards larger and fewer ice clusters faster (e.g., LWC$_0=0$ and LWC$_0=0.15$ at $t=\SI{10}{\second}$ in Fig.~\ref{fig:7}A). While this accelerated coarsening in partially wet snow is visually apparent, plot of the normalized interface length $\overline{S}_a$ does not readily capture this difference (Fig.~\ref{fig:7}B, black and red lines). Note that $\overline{S}_a$ in the partially wet snow accounts for the interface of the combined phase ice and water $(\phi_i+\phi_w)$. Meanwhile, the coarsening speed of the fully wet snow is much faster, as captured visually in Fig.~\ref{fig:7}A and also in terms of the normalized ice interface length $\overline{S}_i$ (Fig.~\ref{fig:7}B, blue line).

Wet snow also experiences a broader range of dynamic temperatures than dry snow during metamorphism (Fig.~\ref{fig:8}). This may be explained by the fact that in wet snow, the thermal fluctuations are dominated by the latent heat released/absorbed during freezing/melting process, which is orders of magnitude faster than the process of deposition/sublimation in dry snow (see Table~\ref{tab:kinetic_constr} in the Supporting Information). The temperature distribution also reflects the Gibbs-Thomson effect: because the equilibrium temperature ($T^\rm{equil}$, Eq.~\eqref{eq:Tequil}) decreases with increasing ice grain curvature, smaller ice grains require colder temperature ($T^\rm{equil}<T_\rm{melt}$) to be stable (Fig.~\ref{fig:8}). Because we impose a temperature $T_\rm{melt}$ at the boundaries, which is higher than the equilibrium temperature for curved grains, we find that all ice grains eventually melt away in the case of fully saturated snow (Figs.~\ref{fig:7}A, C, LWC$=0.46$). Note that here, if we initialize the simulation with \emph{concave} grain shapes, this will lead to $T^\rm{equil}>T_\rm{melt}$ and complete freezing of the water in the case of fully wet snow. 

A key characteristic of dry snow metamorphism is the reduction in microstructure surface area while maintaining an almost constant ice volume (Fig.~\ref{fig:7}B-C, black curves). Interestingly, we find that for the partially wet snow case investigated here, the volume of the air phase is also roughly constant throughout the simulation (not shown here). We attribute this to the fact that water storage capacity of the air is limited (the initial vapor concentration is already close to the saturated vapor density). Nevertheless, we do observe a small mass exchange between the ice and water phases for partially wet snow, as evidenced by the changes in $V_i$ for LWC$_0=0.15$ in Fig.~\ref{fig:7}C.

\section{Numerical challenges and model improvements}

%\bl{Should we mention the model should not be used for equilibrium scenarios??}
%\red{I think we should. Could you provide a brief summary of the issue here? I will then edit.}

We have so far presented a series of 1D and 2D simulations of wet snow in domains of limited sizes ($\sim 0.2$mm). The reason we do not explore larger problems in 3D is due to the high computational cost associated with these simulations. This limits our ability to compare with experimental studies, which are always in 3D and done on samples that are $\sim$cm in size (e.g., in \citet{colbeck1986statistics}). Here we summarize these computational challenges as well as points for model improvements in future work.

In order to accurately capture the kinetics defined by the Gibbs-Thomson conditions, our model requires that the numerical interface width parameter $\varepsilon<10^{-8}\,$m (see Section~\ref{sec:kinet_param}). The consequence of this requirement is the need for a rather fine numerical mesh that resolves such interface, making it computationally expensive to simulate larger problems. An adaptive mesh refinement algorithm, rather than the uniform mesh used in this work, would allow one to run mm- and cm-scale simulations in reasonable computational times. Additionally, mathematical additions to our model based on an asymptotic analysis for the thermal problem with asymmetric diffusivities could further improve robustness of our model at the interface.

Another cause for high computational cost is the need to resolve the kinetics of different processes. In this work, we resort to a monolithic scheme, where our time step size is limited by the fastest process. We have alleviated this issue partially using the time-scaling strategy explained in Section~\ref{sec:time_scaling}. Another way to speed up time integration is to neglect the phase transitions related to the vapor phase during melting/freezing scenarios. The errors introduced with this approach would be small as long as melting/freezing occurs for short time intervals compared to the total time of interest.

Finally, our model is currently not suited for studying the \textit{dynamics} of the triple junction ---the region where the air, liquid and ice phases contact--- when the system is near the triple point. %This is because our phase-field formulation accounts for surface tension and the mobility is not constant (Eq.~\eqref{eq:mobility}). 
At the triple point, defined here as $\{T,\rho_v\}=\{T_\rm{melt},\rho^I_{vs}(T_\rm{melt})\}$, the triple junction is expected to display an equilibrium configuration defined by the surface tensions in Eq.~\eqref{eq:surf_tens_eta}. In our model, an initially non-equilibrium triple junction configuration will evolve towards equilibrium through the co-movement of the water-air, air-ice and water-ice interfaces. However, because each interface moves at different speed, as defined by the phase-dependent mobility (Eq.~\eqref{eq:mobility}), a true equilibrium may not be possible. Instead, the model produces oscillations around the triple junction that impede the achievement of the equilibrium configuration \cite{miyoshi2016validation,moelans2022new}. 
Far away from the triple point, phase change kinetics around the triple junction dominate over the surface tension kinetics and these oscillations do not appear.

For future work, we will focus on more realistic representation of the nucleation process that includes randomized Gaussian noise \cite{karma1999phase,cates2023classical}, which would produce a more realistic behavior of the wet snow metamorphism problem. The current model also does not consider the density difference between the solid ice and liquid water phase, which would introduce small mass conservation error (the model is volume conserved). Our model might be expanded based on \citet{huang2022consistent,zhang2022phase,hagiwara2017ice} to account for mass conservation during melting/freezing, but further research is needed.

\section{Conclusions}
We propose a non-variational phase-field model for wet snow metamorphism.
The model accounts for the ice, water, and air phases, along with the temperature and vapor dynamics, and captures the actual kinetics of solidification, sublimation, and evaporation (and the opposite transitions). 
To the best of our knowledge, this is the first phase-field model that simultaneously reproduces the different phase transitions of water.
The model results unveil the intimate coupling amongst the various transport and phase change processes involved in wet snow metamorphism. Our results show that local humidity conditions affect snow melting and melt refreezing.
The results also reveal the differences between dry and wet snow metamorphism when water flow is not considered: partially and fully wet snow experience larger thermal fluctuations due to the dominant freeze/melt processes and the coarsening rate is higher for fully wet snow compared to dry snow as expected. However, our observations are based on numerically generated partial wet snow where a liquid film of constant thickness is imposed over all ice grain surfaces. We recognize that this way of initializing the simulation may not represent actual liquid film distribution in wet snow and thus may result in unrealistic dynamics. In addition, because we do not consider thin film flow and other capillary-driven processes of the liquid phase, our model may be missing key fluid mechanical processes that influence how wet snow evolves. 

The current results focus on 1D and 2D problems, due to the high computational cost of the model equations. Future modification, extension and application of this model in 3D may be used to quantitatively analyze the evolution of the snow pore structure during wet snow metamorphism, which dictates the thermo-mechanical and hydraulic properties of the snow at larger scales. The proposed modeling framework has proven useful to study wet snow metamorphism and may be the basis to investigate other mm- or cm-scale problems involving water phase transitions such as, e.g., the freezing of a water droplet on a surface \cite{snoeijer2012pointy,zeng2022influence,seguy2023role,sebilleau2021air}, the human-induced thermo-mechanical changes of icy planetary surfaces, or water spray cooling.

%While the literature reviewed here focus on snow microstructure, the interplay between ice, liquid water, and vapor at the microscopic scale ($<$mm) is also relevant to many other problems in engineering such as the Leidenfrost effect \cite{edalatpour2022three} and thermal spray coating \cite{herbaut2019liquid,park2021using}, airplane icing \cite{chu2019bubble,seguy2023role}, or anti-icing and anti-frosting applications \cite{witt2019ice,hauer2021frost}.

\begin{acknowledgement}
The authors acknowledge the partial support from the Resnick Sustainability Institute at California Institute of Technology, the National Science Foundation under Grant No. EAR-2243631 and the ACS Petroleum Research Fund Doctoral New Investigator Grant No. 66867-DNI9. The authors also acknowledge insightful discussions with Dr. Quirine Krol.
\end{acknowledgement}
\bibliography{snow_metamorphism}
\clearpage

\begin{suppinfo}
\setcounter{section}{0}
\section{Equivalence to models of two-phase transition}\label{sec:2PMequivalence}

In this section we show that our wet snow metamorphism model is equivalent to the two-phase models for solidification, sublimation, and evaporation when only two phases are present in the system. 
Only for this section, we assume that $\theta_\rm{nucl}=0$ (see Eqs.~\eqref{eq:ice_simpl}--\eqref{eq:air_simpl} in the main text).
To prove the equivalency, we first need to show that a two-phase system remains as a two-phase system and the non-existing phase does not appear in $\Omega$.
It is trivial to show that the phase-fields equations in our model, Eqs.~\eqref{eq:ice_simpl}--\eqref{eq:air_simpl} in the main text, have the property $\phi_l(\bsx,t)=0\;\forall t$ if $\phi_l(\bsx,0)=0$ and $\theta_\rm{nucl}=0$, which proves this first condition.
The second condition constitutes the actual equivalence, which we analyze for each phase transition in the following paragraphs.

\subsection{Solidification}

Solidification and melting occurs when $\phi_a=0$.
We can rewrite the $\phi_i$ evolution equation (Eq.~\eqref{eq:ice_simpl} in the main text) imposing $\phi_a=0$ as
\begin{equation}\label{eq:ice_cond2}
    \frac{\partial \phi_i}{\partial t} = -3 M_0\left( \frac{1}{\varepsilon} \phi_i(1-\phi_i)(1-2\phi_i)   - \varepsilon \nabla^2\phi_i \right) - \alpha_\rm{sol}\phi_i^2\phi_w^2\frac{T-T_\rm{melt}}{L_\rm{sol}/c_{p,w}}.
\end{equation}
The Lagrange multiplier in our formulation guarantees that $\displaystyle\phi_i+\phi_w=0$, which implies that $\displaystyle\frac{\partial\phi_w}{\partial t}=-\frac{\partial\phi_i}{\partial t}$.
%We could also prove this point by imposing $\phi_a=0$ into Eq.~\eqref{eq:wat_simpl}. 
Imposing $\phi_a=0$, the temperature evolution equation (Eq.~\eqref{eq:tem_simpl} in the main text) is expressed as:
\begin{equation}\label{eq:tem_cond2}
    \rho(\overline{\phi})c_p(\overline{\phi})\frac{\partial T}{\partial t} = \div\left[ K(\overline{\phi})\grad T \right] + \rho(\overline{\phi})L_\rm{sol}\frac{\partial \phi_i}{\partial t}.
\end{equation}
We can disregard the vapor density equation since there is no air phase.
Equations~\eqref{eq:ice_cond2} and \eqref{eq:tem_cond2} constitute a two-phase model for solidification (and melting) known as the generalized Stefan problem \cite{karma1996phase,gomez2019review}. 
Under certain parameter regimes, this model reproduces the kinetics defined by the Gibbs-Thomson condition for the ice-liquid water interface, which reads
\begin{equation}\label{eq:SI_GTSOL}
    \frac{T-T_\rm{melt}}{L_\rm{sol}/c_{p,w}}=-d_\rm{sol}\chi-\beta_\rm{sol}v_n,
\end{equation}
where $T$ is the interface temperature, $\beta_\rm{sol}$ is the kinetic attachment coefficient, $d_\rm{sol}$ is the capillary length, $\chi$ is the curvature of the interface (positive for spherical ice grains), and $v_n$ is the normal velocity of the interface (positive for ice growth). Karma and Rappel \cite{karma1996phase,karma1998quantitative} showed that the two-phase model captures the Gibbs-Thomson condition kinetics when (i) $\varepsilon\rightarrow 0$ and (ii) the model parameters $M_0$ and $\alpha_\rm{sol}$ follow the relations
\begin{equation}\label{eq:GT_sol1}
    M_0=M_\rm{sol}=\frac{\varepsilon}{3\tau_\rm{sol}} , \quad
    \alpha_\rm{sol} = \frac{\lambda_\rm{sol}}{\tau_\rm{sol}}, \quad
    d_\rm{sol}=a_1\frac{\varepsilon}{\lambda_\rm{sol}}, \quad \beta_\rm{sol}=a_1\left(\frac{\tau_\rm{sol}}{\varepsilon\lambda_\rm{sol}} - a_2\frac{\varepsilon}{D^*_{iw}} \right),
\end{equation}
where $a_1\approx 5$, $a_2\approx0.1581$ \footnote{The $a_1$ and $a_2$ values are computed by following the same procedure as in \citet{karma1998quantitative}, applied to Eqs.~\eqref{eq:ice_cond2} and \eqref{eq:tem_cond2}.}, and $D^*_{iw}\approx (D_i+D_w)/2$ represents thermal diffusivity, such that $D_j=K_j/(\rho_j c_{p,j})$ for $j=\{i,w\}$. 
However, we note that the above relations assume equal thermal diffusivity around the freezing/melting interface, which is not true for ice and water in our case. Asymptotic analysis assuming asymmetric diffusivities have been conducted for compositional problems \cite{karma2001, toth2015}, but the same type of analysis for the thermal problems remain to be developed. Nevertheless, \citet{kaempfer2009phase} has shown that the phase-field solidification model recovers the sharp-interface solution if $\displaystyle\varepsilon~<~\beta_\text{sol}{D^*_{iw}}\approx 10^{-5}$m, which is true in the simulations presented in this paper.

\subsection{Sublimation}

Sublimation and deposition occurs when $\phi_w=0$.
After imposing $\phi_w=0$, Eqs.~\eqref{eq:ice_simpl}, \eqref{eq:tem_simpl}, and \eqref{eq:vap_simpl} in the main text are expressed as
\begin{align}
    &\frac{\partial\phi_i}{\partial t} = -3 M_0\left( \frac{1}{\varepsilon} \phi_i(1-\phi_i)(1-2\phi_i)   - \varepsilon \nabla^2\phi_i \right) + \alpha_\rm{sub}\phi_i^2\phi_a^2\frac{\rho_v-\rho_{vs}^I(T)}{\rho_i}\label{eq:cond2_sublice}\\
    &\rho(\overline{\phi})c_p(\overline{\phi})\frac{\partial T}{\partial t} = \div\left[ K(\overline{\phi})\grad T \right]  -\rho(\overline{\phi})L_\rm{sub}\frac{\partial \phi_a}{\partial t},\\
    &\frac{\partial(\phi_a\rho_v)}{\partial t}=\div\left[ \phi_a D_v(T)\grad \rho_v \right] + \rho_\rm{SE} \frac{\partial \phi_a}{\partial t},\label{eq:cond2_sublvap}
\end{align}
with $\partial\phi_a/\partial t=-\partial\phi_i/\partial t$. 
These equations constitute the two-phase sublimation model \cite{kaempfer2009phase} and capture the kinetics defined by the Gibbs-Thomson condition
\begin{equation}\label{eq:SI_GTSUB}
    \frac{\rho_v-\rho_{vs}^I(T)}{\rho_{vs}^I(T)}=d_\rm{sub}\chi+\beta_\rm{sub}v_n,
\end{equation}
as long as the parameters $M_0$, $\alpha_\rm{sub}$, and $\rho_\rm{SE}$ follow the relations \cite{kaempfer2009phase}:
\begin{align}\label{eq:GT_sub1}
    &M_0=M_\rm{sub}=\frac{\varepsilon}{3\tau_\rm{sub}} , \quad
    \alpha_\rm{sub} = \frac{\lambda_\rm{sub}}{\tau_\rm{sub}}, \quad \rho_\rm{SE}=\rho_i,\nonumber\\
    &d_\rm{sub}\frac{\rho_{vs}^I}{\rho_i}=a_1\frac{\varepsilon}{\lambda_\rm{sub}}, \quad
    \beta_\rm{sub}\frac{\rho_{vs}^I}{\rho_i}=a_1\left(\frac{\tau_\rm{sub}}{\varepsilon\lambda_\rm{sub}} - a_2\frac{\varepsilon}{D^*_{ia}}- a_2\frac{\varepsilon}{D_{v0}} \right),
\end{align}
where $\beta_\rm{sub}$ and $d_\rm{sub}$ are the Gibbs-Thomson coefficients for sublimation and $D^*_{ia}$ represents the average thermal diffusivity of ice and air. Again, we note that the above relations assume equal thermal diffusivity around the interface, which is not true for ice and air in our case. Asymptotic analysis assuming for this exact problem remains to be developed. Meanwhile, \citet{kaempfer2009phase} has shown that the above phase-field model recovers the sharp-interface solution if $\displaystyle\varepsilon~<~\beta_\text{sub}\frac{\rho_{vs}^I}{\rho_i}{D^*_{ia}}$ and $\displaystyle\varepsilon~<~\beta_\text{sub}\frac{\rho_{vs}^I}{\rho_i}{D_{v0}}$. These two conditions yield the requirement that $\displaystyle\varepsilon~< 10^{-7}$m, which we do not satisfy in the simulations presented. Thus, our numerical results do not strictly recover the Gibbs-Thomson condition for air-ice interfaces.

\subsection{Evaporation}

The derivation for the evaporation case $(\phi_i=0)$ is analogous to the sublimation case. The final two-phase model for evaporation (without flow) is analogous to the 2-phase sublimation model (Eqs.~\eqref{eq:cond2_sublice}--\eqref{eq:cond2_sublvap}), where the evaporation latent heat is expressed as $(L_\rm{eva}=L_\rm{sub}-L_\rm{sol})$. Here, we simply list the final relations between the model parameters $M_0$, $\alpha_\rm{eva}$, $\rho_\rm{SE}$, and the Gibbs-Thomson coefficients  $\beta_\rm{eva}$ and $d_\rm{eva}$, which read
\begin{align}\label{eq:GT_evap1}
    &M_0=M_\rm{eva}=\frac{\varepsilon}{3\tau_\rm{eva}} , \quad
    \alpha_\rm{eva} = \frac{\lambda_\rm{eva}}{\tau_\rm{eva}}, \quad \rho_\rm{SE}=\rho_w,\nonumber\\
    &d_\rm{eva}\frac{\rho_{vs}^W}{\rho_w}=a_1\frac{\varepsilon}{\lambda_\rm{eva}}, \quad 
    \beta_\rm{eva}\frac{\rho_{vs}^W}{\rho_w}=a_1\left(\frac{\tau_\rm{eva}}{\varepsilon\lambda_\rm{eva}} - a_2\frac{\varepsilon}{D^*_{wa}}- a_2\frac{\varepsilon}{D_{v0}} \right),
\end{align}
where $D^*_{wa}$ represents the average thermal diffusivity of water and air. We note again that the above relations assume equal thermal diffusivity around the interface, which is not true for water and air in our case. Asymptotic analysis assuming for this exact problem remains to be developed. Meanwhile, \citet{kaempfer2009phase} has shown that the above phase-field model recovers the sharp-interface solution if $\displaystyle\varepsilon~<~\beta_\text{eva}\frac{\rho_{vs}^W}{\rho_w}{D^*_{wa}}$ and $\displaystyle\varepsilon~<~\beta_\text{eva}\frac{\rho_{vs}^I}{\rho_w}{D_{v0}}$. These two conditions yield the requirement that $\displaystyle\varepsilon~< 10^{-8}$m, which we do not satisfy in the simulations presented. Thus, our numerical results do not strictly recover the Gibbs-Thomson condition for air-water interfaces.

%In conclusion, our model can capture the actual kinetics defined by the Gibbs-Thomson conditions for solidification, sublimation, and evaporation (and opposite transitions) under certain parameter regimes. %To do that, $M_0$ and $\rho_\rm{SE}$ must take different values depending on the phase transition that takes place, i.e., $M_0$ and $\rho_\rm{SE}$ must be phase-dependent functions, which are defined in Section~\ref{sec:phase_dependent_funct} of the main text.
\end{suppinfo}

\section{Additional details for numerical implementation}
\subsection{Determining temporal scaling coefficients}\label{sec:SI_timescale}
Wet snow metamorphism involves processes at different time scales.
The characteristic times for solidification $(t_\rm{sol})$, evaporation $(t_\rm{eva})$, sublimation $(t_\rm{sub})$, thermal diffusion $(t_{T,i},\,t_{T,w},\,\rm{and}\,t_{T,a})$, and vapor diffusion $(t_v)$, assuming a characteristic length scale of $L=10^{-4}\,$m, are
\begin{align} \label{eq:time_scales}
    & t_\rm{sol} \equiv L/v_{n,\rm{sol}}\sim 1\,\rm{s}, \; t_\rm{sub}\equiv L/v_{n,\rm{sub}}\sim 10^{4}\,\rm{s}, \; t_\rm{eva} \equiv L/v_{n,\rm{eva}} \sim 10^{3}\,\rm{s}, \; t_{T,i} \equiv L^2/D_i \sim 10^{-2}\,\rm{s}, \nonumber\\  
    & t_{T,w} \equiv L^2/D_w \sim 10^{-1}\,\rm{s}, \; t_{T,a} \equiv L^2/D_a \sim 10^{-3}\,\rm{s},\; t_v \equiv L^2/D_{v0} \sim 10^{-4}\,\rm{s},
\end{align}
where thermal diffusivities are calculated as $D_l=K_l/(\rho_l c_{p,l})$, for phase $l$, and we consider typical interface velocities $v_{n,j}$, for $j=\{\rm{sol},\rm{sub},\rm{eva}\}$ (see Table~\ref{tab:kinetic_constr}).
These values show that sublimation and evaporation are several orders of magnitude slower than solidification and any diffusion process. 
A monolithic numerical solver requires the use of time steps that capture the lower characteristic time (i.e., $t_v$), which would significantly increase the computational times. 

Here, we leverage the procedure explained in \citet{kaempfer2009phase} to speed up the simulations. 
This approach leverages the fact that $T$ and $\rho_v$ exhibit a quasi-steady state compared to the phase transition kinetics. 
The procedure consists of multiplying the right-hand side of the $T$ and $\rho_v$ equations (Eqs.~\eqref{eq:tem_simpl} and \eqref{eq:vap_simpl} in the main text, respectively) by a time-scale factor, so that the characteristic times for vapor and thermal diffusion increase some orders of magnitude while $T$ and $\rho_v$ keep displaying a quasi-steady state.
This approach enables the use of larger time steps without introducing noticeable errors in the simulations.
We multiply the right-hand side of the $\rho_v$ equation by the time-scale factor $\xi_v<1$, which provokes an increase of the vapor diffusion characteristic time to $t^*_v=t_v/\xi_v$.
As long as $t^*_v$ does not exceed $t_\rm{sol}$, $t_\rm{sub}$, and $t_\rm{eva}$,
%, i.e., as long as $t^*_v<t_\rm{sol}$, 
vapor concentration displays a quasi-steady state and the overall problem kinetics is not affected. This condition leads to $\xi_v>10^{-4}$.
For the $T$ equation we consider the largest thermal diffusion characteristic time, $t_{T,w}$.
In Eq.~\eqref{eq:time_scales} we observe that $t_{T,w}\sim t_\rm{sol}$, while $t_{T,w}< t_\rm{sub}$ and $t_{T,w}< t_\rm{eva}$. Thus, $T$ displays a quasi-steady state when solidification does not occur. 
We follow the same procedure and multiply the right-hand side of the $T$ equation by the time-scale factor $\xi_T<1$ when solidification does not take place. 
As long as the new characteristic time $t^*_{T,w}=t_{T,w}/\xi_T$ does not exceed $t_\rm{sub}$ and $t_\rm{eva}$, temperature displays a quasi-steady state and the problem kinetics is not affected. This condition leads to $\xi_T>10^{-4}$.
In this work, we take $\xi_v=10^{-3}$ and $\xi_T=1$ or $\xi_T=10^{-2}$ depending on whether or not solidification occurs in any parts of the domain.

\subsection{Kinetic parameters constraints on grid resolution}\label{sec:kinet_param_SI}

If we assume equal diffusivity for all phases, then the relations between the Gibbs-Thomson parameters $(\beta_j,d_j)$ and our model parameters $(M_j,\alpha_j)$ (for $j=\{\rm{sol},\rm{sub},\rm{eva}\}$) can be derived from asymptotic analysis \cite{karma1996phase,karma1998quantitative} and are valid under the following conditions:
\begin{equation}\label{eq:kinet_constrain}
    1.\;\chi\varepsilon<1,\quad\quad 2.\;\varepsilon v_{n,j}/D_k<1, \quad\quad 3.\;\lambda_j u_j<1, \quad\quad 4.\; \tau_j v_{n,j} / \varepsilon <1,
\end{equation}
for $k=\{i,w,a,v0\}$, where $\varepsilon$ represents the interface width, $D_k$ is the thermal or vapor diffusivity coefficient, $v_{n,j}$ is the normal velocity of the interface, $\lambda_j$ and $\tau_j$ are parameters defined in Section~\ref{sec:2PMequivalence} in the Supporting Information, and $u_j$ is the phase-change factor that multiplies the terms $\alpha_j\phi_l^2\phi_m^2$ in the phase-field equations (for $\{l,m\}=\{i,w,a\}$; see Eqs.~\eqref{eq:ice_final} and  \eqref{eq:air_final} in the main text).
Condition~1 imposes an interface width smaller than the curvature radius,
condition~2 guarantees that thermal and vapor diffusion are faster than the interface motion, and
conditions~3 and 4 ensure the preservation of a tanh-profile (i.e., quasi-equilibrium profile) across the interface. We use these conditions to guide the choice of $\varepsilon$ in our model; however, we recognize that these conditions are only valid for equal diffusivity problems, which is not true for the wet snow problem.
These conditions impose a restriction on $\varepsilon$, which represents the numerical interface width and is an adjustable model parameter. 

\begin{table}[bt]
    \caption{Gibbs-Thomson parameters and $\varepsilon$ limitations. Index $j$ stands for solidification, sublimation and evaporation.  The $\beta_j$, $d_j$, $M_j$, and $\alpha_j$ values considered in this work are listed in the bottom rows. }
    \label{tab:kinetic_constr}
    \begin{tabular}{lccc}
        \hline\hline
         & Solidification & Sublimation & Evaporation \\
        \hline 
        $\beta_j$ (s/m) & $\sim(10 , 100)$ & $(2 , 200)\times 10^{4}$ & $(2 , 200)\times 10^{3}$ \\
        $d_j$ (m) & $\sim 5\times 10^{-10}$ & $\sim  10^{-9}$ & $\sim 5\times 10^{-10}$ \\
        $u_j$ (-) & $\sim  10^{-3}$ & $\sim  10^{-8}$  & $\sim  10^{-8}$ \\
        $ v_{n,j}$ (m/s) & $\sim ( 10^{-4},10^{-5})$  & $\sim (10^{-7},10^{-9})$ & $\sim  (10^{-6},10^{-8})$ \\
        Cond.~1 (m)  & $\varepsilon<\,\sim 10^{-4}$ & $\varepsilon<\,\sim 10^{-4}$ & $\varepsilon<\,\sim 10^{-4}$ \\
        Cond.~2 (m)  & $\varepsilon<\,\sim 10^{-3}$ & $\varepsilon<\,\sim  10$ & $\varepsilon<\,\sim  10^{-1}$  \\
        Cond.~3 (m)  & $\varepsilon<\,\sim 10^{-7}$ & $\varepsilon<\,\sim 10^{-7}$ & $\varepsilon<\,\sim 10^{-8}$ \\
        Cond.~4 (m)  & $\varepsilon<\,\sim 10^{-7} $ & $\varepsilon<\,\sim 10^{-7}$ & $\varepsilon<\,\sim  10^{-7}$  \\
        \hline
        \multicolumn{4}{c}{Parameter values considered in this work  (lead to $\varepsilon<10^{-6}$)}  \\
        $\beta_j$ (s/m) & $125$ & $1.4\times 10^{5}$ & $1.53\times 10^{4} $ \\
        $d_j$ (m) & $ 4\times 10^{-9}$ & $ 10^{-7}$ & $ 6\times 10^{-8}$ \\
        $M_j$ (m/s) & $2.12\times10^{-5}$ & $4.33\times10^{-7}$ & $1.34\times10^{-6}$ \\
        $\alpha_j$ (1/s) & $7.96\times10^{4}$ & $1.3\times10^{7}$ & $6.69\times10^{7}$ \\
        \hline\hline
    \end{tabular}
\end{table}
To find a valid range for $\varepsilon$, we first estimate the Gibbs-Thomson parameters $\beta_j$ and $d_j$ from the literature \cite{eames1997evaporation,libbrecht2017physical,kaempfer2009phase,krishnamachari1996gibbs} (see Table~\ref{tab:kinetic_constr}). 
Note that $\beta_j$ may range several orders of magnitude depending on ambient conditions and growth type \cite{libbrecht2017physical}.
We next estimate the typical values of $u_j$ and $v_{n,j}$ (Table~\ref{tab:kinetic_constr}), based on $T$ and $\rho_v$ distributions observed during the different phase transitions. 
We use the Gibbs-Thomson conditions defined in Section~\ref{sec:2PMequivalence} of the Supporting Information to relate $u_j$ and $v_{n,j}$, under the assumption that the curvature $\chi\approx0$.
In order to express $\lambda_j$ and $\tau_j$ as a function of $\beta_j$ and $d_j$ (see Eqs.~\eqref{eq:GT_sol1}, \eqref{eq:GT_sub1}, and \eqref{eq:GT_evap1}), we assume $\rho^I_{vs}/\rho_i=\rho^W_{vs}/\rho_w= 5\times10^{-6}$, which is the approximate value at the freezing point.
For condition 1, we consider typical ice curvatures $\chi\sim 10^4\,\rm{m}^{-1}$.
With this information, we solve for $\varepsilon$ in Eq.~\eqref{eq:kinet_constrain}.
We indicate the less favorable valid range of $\varepsilon$ in Table~\ref{tab:kinetic_constr}.
We find that condition~3 is the most restrictive, such that $\varepsilon<10^{-8}\,\rm{m}$. 

In order to speed up the simulations, we consider a value of $d_\rm{sol}$, $d_\rm{eva}$ and $d_\rm{sub}$ one or two orders of magnitude higher than its actual value (see bottom rows of Table~\ref{tab:kinetic_constr}).
This is because the $T$- and $\rho_v$-terms in the Gibbs-Thomson conditions (Eqs.~\eqref{eq:SI_GTSOL} and \eqref{eq:SI_GTSUB}) dominate the interface motion compared to the curvature terms ($d_j$) for $\chi\sim 10^4\,\rm{m}^{-1}$ in most cases.
Doing so, the $\varepsilon$ constraint in Eq.~\eqref{eq:kinet_constrain} relaxes to $\varepsilon<10^{-6}\,\rm{m}$, which allows us to use a coarser spatial discretization, whereas the wet snow metamorphism kinetics is not significantly affected. 
If compared against experiments, deviations in our modeled kinetics from the observed kinetics could be attributed to this numerical treatment and, hence, the actual values of 
$d_\rm{sol}$, $d_\rm{eva}$ and $d_\rm{sub}$ must be considered.

\subsection{Parameter values}

We list the value of the parameters used in our model at the bottom of Table~\ref{tab:kinetic_constr} and in Table~\ref{tab:parameter_val}.
For simplicity, we assume that the values of these parameters are constant.
The values of the parameters $k_j$ and $g_j$ in the saturated vapor pressures are listed in Table~\ref{tab:sat_vapor}.
The parameter $\xi_T$ is defined as
\begin{equation}\label{eq:xiT}
    \xi_T=\left\{\begin{array}{ll}
    1 & \rm{if}\quad \Gamma_{iw} > \epsilon, \\
    10^{-2} & \rm{if}\quad \Gamma_{iw}\leq \epsilon,
    \end{array} \right.
\end{equation}
where $\Gamma_{iw}$ is the length (in 2D) of the ice-water interface and the parameter $\epsilon$ is close to zero. In particular, we compute the interface as $\Gamma_{iw}=\int_\Omega \phi_i^2\phi_w^2 \rm{d}x$ and we take $\epsilon=2\times 10^{-4}\varepsilon$ in 1D and $\epsilon=2\times 10^{-6} \varepsilon\sqrt{S_{\Omega}}$ in 2D, where $S_{\Omega}$ is the area of $\Omega$. 

\begin{table}
    \caption{Saturated vapor pressure of ice and water. Interpolation parameter values \cite{flatau1992polynomial,wexler1977vapor}.}
    \label{tab:sat_vapor}
    \begin{tabular}{lcccc}
    \hline\hline
        $j$ & 0 & 1 & 2 & 3  \\
        \hline
        $k_j$ & $-0.5865\times 10^4$ & $0.2224\times 10^2$ &  $0.1375\times 10^{-1}$ & $-0.3403\times 10^{-4}$  \\
        $g_j$ & $-0.2991\times 10^{4}$ & $-0.6017\times 10^{4}$ &  $0.1888\times 10^{2}$ & $-0.2836\times 10^{-1}$\\ 
        \hline\hline
        $j$ & 4 & 5 & 6 & 7\\
        \hline
        $k_j$ & $0.2697\times 10^{-7}$ & $0.6918$ & - & - \\
        $g_j$& $0.1784\times 10^{-4}$ & $-0.8415\times 10^{-9}$ & $0.4441\times 10^{-12}$ & $0.2859\times 10^{1}$  \\
        \hline
    \end{tabular}
\end{table}

\begin{table}[]
    \caption{Parameter values. The values of $\beta_j$, $d_j$, $M_j$, and $\alpha_j$ (where $j$ stands for sol, sub, and eva) are listed in the bottom of Table~\ref{tab:kinetic_constr}. We take $\varepsilon=2\times10^{-7}\,$m for the simulations shown in Section~4.3.3 in the main text. }
    \label{tab:parameter_val}
    \begin{tabular}{lcc}
        \hline\hline 
        Parameter & Value & Units \\
        \hline 
        $\varepsilon$ & $5\times10^{-7}$ & m \\
        $\Lambda$ & 1 & N/m \\
        $\sigma_{iw}$ & 0.033 & N/m \\
        $\sigma_{ia}$ & 0.109 & N/m \\
        $\sigma_{wa}$ & 0.076 & N/m \\
        $T_\rm{melt}$ & 0 & $^\circ$C \\
        $P_a$ & 101325 & Pa \\
        $\alpha_\rm{nucl}$ & $5\times 10^6$ & 1/m \\
        $T_\rm{nucl}$ & variable & $^\circ$C \\
        $\rho_a$ & 1.341 & Kg/m$^3$ \\
        $\rho_i$ & 917 & Kg/m$^3$ \\
        $\rho_w$ & 1000 & Kg/m$^3$ \\
        $c_{p,a}$ & $1.044\times 10^3$ & J/(Kg$\,^\circ$C) \\
        $c_{p,i}$ & $1.96\times 10^3$ & J/(Kg$\,^\circ$C)  \\
        $c_{p,w}$ & $4.2\times 10^3$ & J/(Kg$\,^\circ$C) \\
        $K_a$ & 0.02 & W/(m$\,^\circ$C) \\
        $K_i$ & 2.29 & W/(m$\,^\circ$C) \\
        $K_w$ & 0.554 & W/(m$\,^\circ$C) \\
        $L_\rm{sol}$ & $3.34\times 10^5$ & J/Kg \\
        $L_\rm{sub}$ & $2.83\times 10^6$ & J/Kg \\
        $D_{v0}$ & $2.178\times 10^{-5}$ & m$^2$/s \\
        $\xi_v$ & $10^{-3}$ & (-) \\
        $\xi_T$ & see Eq.~\eqref{eq:xiT} & (-) \\
        \hline
    \end{tabular}
\end{table}

\subsection{Initial and boundary conditions}

For the phase fields and the temperature we consider the boundary conditions
\begin{equation}\label{eq:BC1}
    \grad\phi_i\cdot\bsn=0, \quad \grad\phi_a\cdot\bsn=0,\quad T=T_B(\bsx,t)\quad \rm{on}\;\partial\Omega,
\end{equation}
where $\partial\Omega$ is the boundary of $\Omega$ and the boundary temperature $T_B$ may take different values on each boundary and may evolve in time.
For the vapor concentration, we consider a fixed concentration on the boundary or we impose no-flux through the boundary, depending on each simulation, which are expressed as
\begin{equation}\label{eq:BC2}
    \rho_v=\rho_{v,B}\quad\rm{or}\quad \grad\rho_v\cdot\bsn=0\quad \rm{on}\;\partial\Omega,
\end{equation}
where $\rho_{v,B}$ is a constant value.
The initial conditions of our problem may be written as
\begin{equation}
    \phi_l(\bsx,0)=0.5\left(1+\rm{tanh}\left(\frac{d_{l}(\bsx)}{2\varepsilon}\right)\right),\quad T(\bsx,0)=T_0(\bsx), \quad \rho_v(\bsx,0)=\rho^I_{vs}(T_0),
\end{equation}
where $l=\{i,a\}$, $d_{l}$ is the signed distance to the interface of phase $\phi_l$ and $T_0(\bsx)$ is a linear function consistent with the temperature boundary condition $T_B$. 
The distance $d_{l}$ depends on the geometry considered in each simulation.

\subsection{Spatial, time discretization and variable regularization}\label{sec:discretization}

We use the Finite Element Method to solve our model. In particular, we use a uniform mesh composed of bilinear (linear in 1D) basis functions. 
%with bilinear (linear in 1D) basis functions.
We compute the weak form by multiplying the model equations (Eqs.~\eqref{eq:ice_final}--\eqref{eq:air_final} in the main text) by weighting functions, integrating over $\Omega$, and integrating by parts taking into account the boundary conditions defined in Eqs.~\eqref{eq:BC1} and \eqref{eq:BC2}.
To obtain the Galerkin form, we substitute the problem unknowns and the weighting functions by the basis functions.
The phase field variables may take values $\phi_l<0$ and $\phi_l>1$ during transient states, with $l=\{i,w,a\}$.
To avoid singularities and numerical issues we regularize the functions $\rho$, $c_p$, $K$, and $\rho_\rm{SE}$: 
We restrict the value of $\phi_l$ between 0 and 1 only when we compute the functions $\rho$, $c_p$, and $K$.
For the regularization of $\rho_\rm{SE}$ we
(1) implement an additional cubic interpolation locally around $\phi_a=0$ which smooths out the discontinuity at that point, and
(2) extend $\rho_\rm{SE}$ off the ternary diagram constant in the normal direction to the diagram sides.

\end{document}